\documentclass[superscriptaddress,preprintnumbers,nofootinbib,10pt]{revtex4}
\usepackage[dvipdf,dvips]{graphicx} 
\usepackage{amsmath}
\usepackage{amssymb}
\usepackage{comment}
\usepackage{float}
\usepackage{bm}
\usepackage{tabularx}
\usepackage{color}
\usepackage{xcolor}
\usepackage{array} 
\usepackage[normalem]{ulem}

\usepackage{hyperref}
\hypersetup{
colorlinks=true,
citecolor=blue,
citebordercolor=red,
linktoc=all,
linkcolor=blue,
urlcolor=blue
}

\catcode`\@=11
\def\lsim{\mathrel{\mathpalette\@versim<}}
\def\gsim{\mathrel{\mathpalette\@versim>}}
\def\@versim#1#2{\vcenter{\offinterlineskip
\ialign{$\m@th#1\hfil##\hfil$\crcr#2\crcr\sim\crcr } }}
\catcode`\@=12
\newcommand{\Slash}[1]{{\ooalign{\hfil/\hfil\crcr$#1$}}}

\newcommand{\T}{\text{T}}
\newcommand{\TT}{\text{TT}}

\newcommand{\p}{\partial}

\newcommand{\bp}{\begin{pmatrix}}
\newcommand{\ep}{\end{pmatrix}}
\newcommand{\nn}{\nonumber\\}
\newcommand{\paren}[1]{\left(#1\right)}

\newcommand{\df}{\text{d}}

\newcommand{\bs}[1]{\boldsymbol}

\newcommand{\Tr}{{\rm Tr}\,}
\newcommand{\mat}[1]{\begin{matrix}#1\end{matrix}}
\newcommand{\pmat}[1]{\begin{pmatrix}#1\end{pmatrix}}

\newcommand{\fn}[1]{\!\left(#1\right)}

\usepackage{etoolbox}
\let\bbordermatrix\bordermatrix
\patchcmd{\bbordermatrix}{8.75}{4.75}{}{}
\patchcmd{\bbordermatrix}{\left(}{\left[}{}{}
\patchcmd{\bbordermatrix}{\right)}{\right]}{}{}

\newcolumntype{C}{>{$}c<{$}}

\allowdisplaybreaks

\begin{document}

\begin{flushright}
CCTP-2019-4, ITCP-IPP 2019/4
\end{flushright}

\title{
On the impact of Majorana masses in gravity-matter systems
}

\author{Gustavo P. de Brito}
\email{gpbrito@cbpf.br}
\affiliation{CBPF $-$ Centro Brasileiro de Pesquisas F\'isicas, Rua Dr. Xavier Sigaud 150, 22290-180, Rio de Janeiro, RJ, Brazil}
\affiliation{Institut f\"ur Theoretische Physik, Universit\"at Heidelberg, Philosophenweg 16, 69120 Heidelberg, Germany}

\author{Yuta \surname{Hamada}}
\email{yhamada@physics.uoc.gr}
\affiliation{Crete Center for Theoretical Physics, Institute for Theoretical and Computational Physics, Department of Physics, University of Crete, P.O. Box 2208, 71003 Heraklion, Greece}

\author{Antonio D. \surname{Pereira}}
\email{adpjunior@id.uff.br}
\affiliation{Instituto de F\'isica, Universidade Federal Fluminense, Campus da Praia Vermelha, Av. Litor\^anea s/n, 24210-346, Niter\'oi, RJ, Brazil}
\affiliation{Institut f\"ur Theoretische Physik, Universit\"at Heidelberg, Philosophenweg 16, 69120 Heidelberg, Germany}

\author{Masatoshi \surname{Yamada}}
\email{m.yamada@thphys.uni-heidelberg.de}
\affiliation{Institut f\"ur Theoretische Physik, Universit\"at Heidelberg, Philosophenweg 16, 69120 Heidelberg, Germany}

\begin{abstract}
We investigate the Higgs-Yukawa system with Majorana masses of a fermion within asymptotically safe quantum gravity.
Using the functional renormalization group method we derive the beta functions of the Majorana masses and the Yukawa coupling constant and discuss the possibility of a non-trivial fixed point for the Yukawa coupling constant.
In the gravitational sector we take into account higher derivative terms such as $R^2$ and $R_{\mu\nu}R^{\mu\nu}$ in addition to the Einstein-Hilbert term for our truncation.
For a certain value of the gravitational coupling constants and the Majorana masses, the Yukawa coupling constant has a non-trivial fixed point value and becomes an irrelevant parameter being thus a prediction of the theory.
We also discuss consequences due to the Majorana mass terms to the running of the quartic coupling constant in the scalar sector.
\end{abstract}
\maketitle
\section{Introduction}
The formulation of a consistent theory which fundamentally describes the quantum fluctuations of gravitational degrees of freedom is still an open issue in theoretical physics. Different approaches to this problem based on different theoretical assumptions are being developed and each of them has internal challenges to be handled. Notably, the quantization of General Relativity within the standard perturbative quantum field-theoretic framework leads to a perturbatively non-renormalizable quantum theory. Among several different possibilities to circumvent this issue, Asymptotic Safety is one route which stays within the continuum quantum field theory realm, but goes beyond the standard perturbative paradigm \cite{Hawking:1979ig,Niedermaier:2006wt,Niedermaier:2006ns,Percacci:2007sz,Reuter:2012id,Nagy:2012ef,Codello:2008vh,Percacci:2017fkn,Eichhorn:2018yfc,Reuter:2019byg}. 

In this approach, a fundamental quantum theory of gravity could be realized thanks to the existence of a non-trivial ultraviolet (UV) fixed point in the renormalization group flow. At this fixed point, the theory becomes scale invariant \cite{Wetterich:2019qzx} and one is allowed to zoom in up to arbitrarily short distances without running into divergences. Being an interacting fixed point, the counting of relevant directions in theory space, i.e., the number of free parameters to be fixed by experiments is not controlled by the canonical dimensionality of the corresponding couplings. By integrating out quantum fluctuations, all terms compatible with the symmetries of the underlying theory will be generated in the quantum effective action yielding an infinite-dimensional theory space. For the theory to be predictive, the number of relevant parameters should be finite. Since one looks for an interacting fixed point, its discovery might not be feasible within perturbation theory and new techniques might be necessary. Thus, an asymptotically safe quantum theory of gravity relies on the existence of a non-trivial UV fixed point which features finitely many relevant directions.

The first evidence for the existence of the aforementioned non-trivial fixed point was found with the $\epsilon$-expansion method in $2+\epsilon$ dimensions~\cite{Hawking:1979ig,Kawai:1989yh}.
However, in order to establish the existence of the fixed point in four dimensions, other methods are necessary. One possibility is to regularize the path integral for quantum gravity by lattice methods and systematically look for a suitable fixed point which allows for a continuum limit which resembles our universe, see, e.g., \cite{Ambjorn:2012jv,Laiho:2016nlp}. In the asymptotic safety program, most evidence for the existence of a non-trivial fixed point were obtained with the functional renormalization group (FRG), see \cite{Morris:1998da,Berges:2000ew,Aoki:2000wm,Bagnuls:2000ae,Polonyi:2001se,Pawlowski:2005xe,Gies:2006wv,Delamotte:2007pf,Rosten:2010vm,Braun:2011pp} for reviews. This was pioneered by Reuter in \cite{Reuter:1996cp} where a UV fixed point was found using the FRG within the Einstein--Hilbert truncation for the effective average action, a scale-dependent action which takes into account the integration of quantum fluctuations up to some scale $k$. Several subsequent works, employing more sophisticated approximations provided compelling support for the existence of such the Reuter fixed point, see, e.g., \cite{Souma:1999at,Reuter:2001ag,Litim:2003vp,Codello:2006in,Benedetti:2009rx,Benedetti:2009gn,Manrique:2010am,Manrique:2011jc,Christiansen:2012rx,Falls:2013bv,Benedetti:2013jk,Codello:2013fpa,Falls:2014tra,Christiansen:2014raa,Christiansen:2015rva,Gies:2015tca,Gies:2016con,Biemans:2016rvp,Christiansen:2016sjn,Denz:2016qks,Knorr:2017fus,Knorr:2017mhu,Christiansen:2017bsy,Falls:2017lst,Falls:2018ylp}. Remarkably, the existence of the UV fixed point has shown to be preserved against the introduction of Standard Model (SM) matter degrees of freedom, hinting to a fundamental theory of quantum gravity which is compatible with the observed SM particles, see \cite{Percacci:2002ie,Percacci:2003jz,Narain:2009fy,Zanusso:2009bs,Eichhorn:2011pc,Eichhorn:2012va,Dona:2013qba,Dona:2014pla,Labus:2015ska,Oda:2015sma,Meibohm:2015twa,Dona:2015tnf,Meibohm:2016mkp,Eichhorn:2016esv,Eichhorn:2016vvy,Biemans:2017zca,Hamada:2017rvn,Christiansen:2017qca,Eichhorn:2017eht,Eichhorn:2017egq,Eichhorn:2017sok,Christiansen:2017cxa,Eichhorn:2017als,Alkofer:2018fxj,Eichhorn:2018akn,Eichhorn:2018ydy,Eichhorn:2018yfc,Eichhorn:2018nda,Pawlowski:2018ixd}.
Moreover, many different works have provided evidence for a saturation on the number of relevant directions associated to the Reuter fixed point, ensuring thus, predictivity~\cite{Codello:2007bd,Machado:2007ea,Benedetti:2009rx,Benedetti:2009gn,Falls:2013bv,Falls:2014tra,Gies:2016con,Christiansen:2016sjn,Denz:2016qks,Hamada:2017rvn,Falls:2017lst,Falls:2018ylp,deBrito:2018jxt}. This underpins the quest for a quantum theory of gravity by using continuum quantum field theory methods.

Within the asymptotic safety scenario, coupling matter degrees of freedom to (quantum) gravity is straightforward. This allows for a rich interplay between the effects of matter fluctuations into gravitational running couplings and the reverse. In particular, this opens the possibility to test whether quantum gravitational effects allow for the resolution of long-standing problems as, e.g., the triviality problem in the SM, see \cite{Harst:2011zx,Christiansen:2017gtg,Eichhorn:2017lry}. Having a fundamental theory of gravity and matter valid up to arbitrarily short distances also leads to an enhancement of predictivity. In fact, the fixed point structure as well as the assumption that the SM holds up all the way until the Planck scale leads to a prediction of the Higgs mass to be $m_H = 126$\,GeV for the top quark mass $m_t=171$\,GeV, see~\cite{Shaposhnikov:2009pv,Pawlowski:2018ixd}. This happens thanks to its quantum-gravity induced irrelevance, being thus a prediction and not a free parameter of the theory. Recent works~\cite{Eichhorn:2017lry,Eichhorn:2017muy,Eichhorn:2018whv} have managed to successfully explain observed quantities in the low energy regimes by assuming the existence of the asymptotically safe fixed point for gravity-matter systems.
Furthermore, asymptotically safe quantum gravity might be able to provide solutions for important problems in both particle physics and cosmology such as the gauge hierarchy problem~\cite{Wetterich:2016uxm} and the cosmological constant problem~\cite{Wetterich:2017ixo}.

So far, the SM has shown to be consistent with results from the collider experiments up to the TeV scale, and no signal of new physics has appeared. This might be taken as a strong indication that, possibly, the SM is not drastically modified up to very high energy scales as the Planck scale.
However, the SM does not account for dark matter. Many different models to extend the SM consistently with dark matter observations were developed in the last years. Very much like previously discussed, one can investigate if quantum-gravity effects play a relevant role in the description of dark matter phenomenology, see, e.g.~\cite{Eichhorn:2017als}. This allows for low-energy consistency checks of the asymptotic safety scenario.

Another very important issue which is not accommodated in the SM is the (very tiny) neutrino mass. A standing out mechanism which describes the tiny mass for the left-handed neutrino while explains why a right-handed neutrino although necessary to be existent is not easily detected is the so-called seesaw mechanism~\cite{Minkowski:1977sc,GellMann:1980vs,Yanagida:1980xy,Mohapatra:1979ia}. Such a mechanism is grounded on the introduction of a right-handed neutrino with a Majorana mass term which couples to the left-handed neutrino and the Higgs through a Yukawa interaction. Schematically, after spontaneous symmetry breaking, a Dirac mass term is generated and the balance between the Majorana mass parameter and the Dirac term allows for the assignment of a very large mass to the right-handed sector while the left-handed is very light.

In the present work, we aim at giving the first steps to investigate the quantum-gravity fluctuations effects to the seesaw mechanism. To pave the way for such understanding, we consider a toy Higgs-Yukawa system with one scalar and one Dirac fermion coupled to gravity. Different Majorana mass terms are introduced for the right- and left-handed components of the fermion. On top of that, right- and left-handed components interact with the scalar field via a Yukawa coupling. Beta functions for the Majorana masses and Yukawa coupling are computed taking into account quantum gravity effects with the FRG, which, within our approximation, also include contributions coming for curvature squared terms such as $R^2$ and $R_{\mu\nu}R^{\mu\nu}$. We mention that curvature squared terms are important in the framework of perturbative quantum gravity, in such a case, the higher-derivative contributions coming from $R^2$ and $R_{\mu\nu}R^{\mu\nu}$ improve the UV behavior of the propagator, leading to a perturbative renormalizable theory~\cite{Stelle:1976gc}. However, the price to be paid is the appearance of ghost-like states in the perturbative spectrum and, as consequence, unitarity violation~\cite{Stelle:1976gc}, but see also recent discussion on this issue~\cite{Holdom:2016xfn,Donoghue:2017fvm,Anselmi:2018ibi}. From the point of view of FRG, the situation may be quite different. In this case the curvature squared terms are included as part of an approximated (truncated) solution for the FRG equation. As a truncated solution, we cannot perform a reliable analysis of the spectrum of the theory and, as a consequence, the ghost problem remains as an open question in this framework. For a recent discussion on unitarity problem in FRG see, e.g., \cite{Becker:2017tcx}

Assuming the existence of a non-trivial fixed point for the gravitational couplings, we investigate the consequences to the masses and Yukawa coupling.
In particular, regarding the critical exponent of the Yukawa coupling as a function of the Majorana masses and gravitational couplings we investigate the dependence of its sign on these parameters.
We classify possible scenarios in low energy regimes as a consequence of asymptotically safe gravity.
The impact of a non-trivial fixed point of the Yukawa coupling on the quartic scalar coupling is also discussed.
Assuming that the system analyzed in this paper is associated with a simple toy model of the SM, the quartic scalar coupling at the Planck scale receives a finite correction from a fermion with the Majorana masses.
This means that the prediction for the central value of the Higgs mass at the electroweak scale could be modified.
The present work might be viewed as a stepping stone towards the more realistic system which takes into account all the SM properties.

This paper is organized as follow:
In Section~\ref{Introduction to method and model}, we introduce the key concepts of the FRG and the toy model that we will investigate.
Our results are collected in Section~\ref{result section}.
Section~\ref{summary section} is dedicated to discussions and perspectives.
Conventions and computational details are presented in the appendices.

\section{Method and Model}\label{Introduction to method and model}
\subsection{Functional renormalization group}
In order to extract the effects of quantum-gravity fluctuations to the running of masses and Yukawa coupling (as well as the impact of matter fluctuations to gravitational couplings), we use the FRG. 
Its central object is the effective average action $\Gamma_k$. The parameter $k$ acts as an infrared cutoff and $\Gamma_k$ is obtained by integrating out modes with momenta larger than $k$. Lowering $k$ corresponds to integrate more and more momentum shells in the Wilsonian sense.
The change of $\Gamma_k$ with  $k$ is described by the Wetterich equation~\cite{Wetterich:1992yh}, 
\begin{align}
\p_t\Gamma_k =\frac{1}{2}\text{STr}\left[ \left( \Gamma_k^{(2)}+ \textbf{R}_k \right)^{\!-1} \!\! \p_t \textbf{R}_k\right],
\label{Wetterich equation}
\end{align}
where $\p_t$ denotes the derivative with respect to $t=\log\fn{k/k_0}$ (with a reference cutoff $k_0$), $\textbf{R}_k$ is a cutoff function, which suppress all modes with momenta smaller than $k$, $(\Gamma_k^{(2)}+ \textbf{R}_k)^{-1}$ is the regularized full propagator, and $\text{STr}$ stands for the supertrace acting on all internal as well as spacetime indices and carries an appropriate numerical factor depending on the nature of the field.
The Wetterich equation \eqref{Wetterich equation} has a one-loop exact form, a fact that greatly simplifies practical calculations. For some reviews about the Wetterich equation, we refer to \cite{Morris:1998da,Berges:2000ew,Aoki:2000wm,Bagnuls:2000ae,Polonyi:2001se,Pawlowski:2005xe,Gies:2006wv,Delamotte:2007pf,Rosten:2010vm,Braun:2011pp}.

In general, the effective action $\Gamma_k$ is expanded as, 
\begin{align}
\Gamma_k = \sum_i g_i (k)\mathcal{O}_i (\phi)\,,
\label{eaaexp}
\end{align}
where $g_i$ are the dimensionful coupling constants and $\mathcal{O}_i$ denote operators of the local fields $\phi$.
Using the Wetterich equation and expanding its right-hand side on the same $\left\{ \mathcal{O}_i\right\}$ basis, one 
is able to read off the beta functions for the couplings. 
They have the following structure,
\begin{align}
\p_t \tilde g_i=\beta_i\fn{\tilde g}=-d_{i}\tilde g_i+F_i\fn{\tilde g},
\label{rg equations}
\end{align}
where $\tilde g_i=k^{-d_{i}}g_i$ is the dimensionless coupling constant and $d_i$ is the canonical dimension of $g_i$.
The first term on the right-hand side of the equation \eqref{rg equations} is the canonical scaling term, associated with the dimensionality of the coupling and $F_i\fn{g}$ corresponds to quantum corrections. 
Here, let us assume that the system has a fixed point $\tilde g^* = \left\{\tilde{g}^\ast_1,\tilde{g}^\ast_2,\ldots\right\}$ at which the beta functions vanish: $\beta_i\fn{\tilde g^*}=0$ for all $i$.
We linearize the RG equation \eqref{rg equations} around the fixed point,
\begin{align}
\p_t \tilde g_i\simeq \sum_{j}\frac{\p \beta_i\fn{\tilde g}}{\p \tilde g_j}\bigg|_{\tilde g=\tilde g^*}(\tilde g_j-\tilde g_j)=: -\sum_{j}T_{ij}(\tilde g_j-\tilde g_j).
\end{align}
These equations can be diagonalized and trivially solved.
The eigenvalues of $T$, here denoted by $\theta_i$, correspond to the critical exponents associated to the fixed point.
In our notation, positive $\theta_i$ are relevant directions in the RG sense.
The number of relevant directions dictates how many free parameters the theory features.
That is, a theory with less relevant directions is more predictive.
For the Gaussian (trivial) fixed point $\tilde g^*=0$ the critical exponents are given by $\theta_i\simeq d_i$ and the coupling constants with $d_i>0$, i.e., the power-counting renormalizable ones, are relevant.
In contrast, if the theory has a non-trivial fixed point $\tilde g^*\neq 0$, the critical exponents are $\theta_i\simeq d_i-\p F/\p \tilde g_i|_{\tilde g=\tilde g^*}$.
We see that the contribution from the loop correction $\p F/\p \tilde g_i|_{\tilde g=\tilde g^*}$  typically does not vanish at a non-trivial fixed point leading to a non-canonical scaling. Such non-trivial dynamics can drive canonically irrelevant/relevant coupling into relevant/irrelevant ones. In the case where relevant couplings turn to irrelevant ones, the predictivity power of the theory is enhanced.

\subsection{Setting the truncation}
Our goal is to compute the system of beta functions for a Higgs-Yukawa model with a Majorana mass term coupled to gravity. In order to apply the FRG, we make use of the background field method.
In practice, we linearly split the metric\footnote{
Different choices of field parametrization, e.g., exponential splitting $g_{\mu\nu} = \bar{g}_{\mu\alpha} [e^h]^\alpha_{\,\,\,\nu}$, lead to different results for the beta functions. For a discussion of different choices for the parametrization of the quantum fluctuations and their impact on the fixed point structure in pure-gravity systems, see \cite{Nink:2014yya,Gies:2015tca,Ohta:2016npm,Ohta:2016jvw,Ohta:2018sze,deBrito:2018jxt}. For gravity-matter systems, investigations on the choice of parametrization has been performed in \cite{deBrito:2019umw}
}
 into a background metric ${\bar g}_{\mu\nu}$ and fluctuation $h_{\mu\nu}$,
\begin{align}
g_{\mu\nu}={\bar g}_{\mu\nu} + h_{\mu\nu} \,.
\end{align}
Hereafter indices are raised and lowered by the background (inverse) metric.

We investigate the following truncated effective average action,
\begin{align}
\quad \Gamma_k=\Gamma_k^\text{gravity}+\Gamma_k^\text{matter} \,,
\label{general action}
\end{align}
where the gravity sector is given by
\begin{align}
\Gamma_k^\text{gravity}[g]
=\frac{1}{16\pi G} \int\df^4 x \sqrt{g}\,\big[ 2\Lambda - R + a\, R^2 + b \, R_{\mu\nu} R^{\mu\nu} \big] +S_{\rm gf} + S_{\rm gh} \,.
\label{effective action for gravity}
\end{align}
Here, $G$ and $\Lambda$, respectively, correspond to the Newton and cosmological constants, $a$ and $b$ denote higher curvature couplings, and $S_{\rm gf}$ and $S_{\rm gh}$ are the gauge fixing and Faddeev-Popov ghosts actions\footnote{We have alaready restricted the Faddeev-Popov operator to the background metric. This is enough for our purposes in this work.} written as
\begin{subequations}
\begin{align}
S_{\rm gf}	&=	\frac{1}{32\pi \alpha\, G}\int \df^4 x\sqrt{\bar g}\,
{\bar g}^{\mu \nu}F_\mu F_{\nu},
\label{gaugefixedaction} 
\end{align}
\begin{align}
S_{\rm gh}	&=	-\int\df^4 x\sqrt{\bar g}\,\bar C_\mu\left[ {\bar g}^{\mu\nu}\bar\Box+\frac{1-\beta}{2}{\bar \nabla}^\mu{\bar \nabla}^{\nu}		+{\bar R}^{\mu\nu}\right] C_{\nu}, \label{ghostaction}
\end{align}
\end{subequations}
with $F_\mu	:= {\bar \nabla}^\nu h_{\nu \mu}-\frac{\beta +1}{4}{\bar \nabla}_\mu h$ being the standard gauge condition.
Here, $\bar\nabla_\mu$ denotes the covariant derivative with respect to the background, $\bar\Box=\bar \nabla^2$ is the d'Alembertian. In addition, $h:={\bar g}_{\mu\nu}h^{\mu\nu}$ is the trace of $h_{\mu\nu}$; $C_\mu$ and $\bar C_\mu$ are the ghost and anti-ghost fields, respectively; and $\alpha$ and $\beta$ are dimensionless gauge parameters. For the matter sector, we consider the following action,
\begin{align}
\Gamma_k^\text{matter}
=	\int\df^4 x \sqrt{g}&\Bigg[ 		
Z_\psi{\bar \psi}\Slash {D} \psi
+\frac{1}{2}\left(m_M^R \bar \psi^c P_R \psi
+m_M^L \bar \psi^c  P_L \psi
+\text{h.c.}\right)
+ y_D\phi {\bar \psi}\psi
+\frac{Z_\phi}{2}(\p_\mu \phi)^2
+V\fn{\phi^2}
\Bigg],
\label{effective action for matter 1}
\end{align}
where $V\fn{\phi^2}$ is the scalar potential with $\mathbb{Z}_2$ $(\phi\to -\phi)$ symmetry and $Z_{\phi,\psi}$ are field renormalization factors of scalar and fermion fields, respectively, and contribute to the beta functions through the anomalous dimensions expressed as
\begin{align}
\eta_{\phi}&=-\frac{\p_t Z_{\phi}}{ Z_{\phi}},&
\eta_{\psi}&=-\frac{\p_t Z_{\psi}}{ Z_{\psi}}.
\label{anomalous dimensions for matters}
\end{align}
Here, $\Slash D=\gamma^\mu D_\mu$ is the Dirac operator and satisfies $-\Slash D^2=-D^2+{\bar R}/4$; $\psi^c$ is the charge conjugated spinor field, namely, $\psi^c=C\bar\psi^T$; $P_L=(1-{\gamma}^5)/2$, $P_R=(1+{\gamma}^5)/2$ are the chiral projection operators. We refer to Appendix~A for further conventions.
The effective action \eqref{general action} has the discrete $\gamma^5$-symmetry, i.e., $\psi\to \gamma^5\psi$, $\bar\psi\to -\bar\psi  \gamma^5$ and $\phi\to -\phi$, and then the Dirac mass term $\bar\psi \psi$ and the Majorana-Yukawa terms $\phi\bar \psi^c P_{R,L}\psi+\text{h.c.}$ are forbidden, whereas the Majorana mass term is not.
Indeed, even if we put $m_M^L=0$ and $m_M^R\neq0$ at some scale, quantum effects induce a finite $m_M^L$ as we will see in \eqref{schematic beta function of majorana mass}.
We also impose the CP invariance which prohibits the term $\phi\,\bar \psi \gamma^5\psi$.
The CP invariance also makes the parameters $m_m^R$ and  $m_m^L$ to be real ones.
It follows that the Hermitian conjugate of the Majorana mass term is $m_M^R \bar \psi P_L \psi^c+m_M^L \bar \psi  P_R \psi^c$.

In this work, since we employ a simple model where a neutral fermion
interacts with a scalar-field singlet, both the left-and right-handed fermion fields can have Majorana mass terms.
Note however that the left-handed neutrino is a component in the lepton-doublet field involving a hypercharged particle in the SM, so that the Majorana mass term for the doublet field cannot be admitted in the SM before the electroweak symmetry breaking.

\subsection{York decomposition}
For the fluctuation field $h_{\mu\nu}$ and the ghost fields, the York decomposition~\cite{York:1973ia} is employed,
\begin{align}
	h_{\mu\nu}	=	h_{\mu\nu}^\TT + \bar\nabla _{\mu}\xi_{\nu}	+\bar\nabla _{\nu}\xi_\mu	
	+\paren{\bar\nabla_\mu \bar\nabla_{\nu}	-\frac{1}{4}\bar g_{\mu\nu} \bar\Box}\sigma
	+\frac{1}{4}\bar g_{\mu\nu}h,
	\label{gravitondecomposition}
\end{align}
and
\begin{align}
	C_\mu	=	C_\mu^\T	+\bar\nabla_\mu  C, \qquad \qquad
	{\bar C}_\mu = {\bar C}_\mu^\T+\bar\nabla_\mu {\bar C},
	\label{ghostdecomposition}
\end{align}
where $h^\TT$ is transverse and traceless (TT) tensor field with spin 2,  i.e., $\bar\nabla^{\mu}h^\TT_{\mu\nu}=0$ and $\bar g^{\mu\nu}h^\TT_{\mu\nu}=0$; $\xi_\mu$ is the transverse vector field with spin 1, i.e., $\bar\nabla^\mu \xi_{\mu}=0$; and $ \sigma$ and $h:={\bar g}^{\mu\nu}h_{\mu\nu}$ are the scalar fields with spin 0, $C$ and ${\bar C}$ are spin-0 scalar ghost fields, and $C_\mu ^\T$ and $\bar C_\mu ^\T$ are spin-1 transverse vector ghost fields that satisfy $\bar\nabla^\mu C^\T_\mu=\bar\nabla^\mu {\bar C}^\T_\mu=0$.

Such decompositions entail the following Jacobian in the path integral measure:
\begin{equation}
J = \left[\mathrm{det}^\prime_{(1)}\left(-\bar\Box-\frac{\bar R}{4}\right)\right]^{1/2}\left\{\mathrm{det}^{\prime\prime}_{(0)}\left[-\bar\Box\left(-\bar\Box-\frac{\bar R}{3}\right)\right]\right\}^{1/2}\left[\mathrm{det}^{\prime}_{(0)}(-\bar\Box)\right]^{-1} ,
\label{Jacobian from York decomposition}
\end{equation} 
where the primes indicate that the first or the first two eigenvalues are removed from the determinant. Although this Jacobian can be absorbed by redefining the fields $\xi_\mu$, $\sigma$, $\bar{C}$ and $C$, here we take these contributions into account in the flow equation  without any field redefinitions.
This can be done by regularizing the Jacobian with the usual prescription $-\bar{\Box} \mapsto P_k(-\bar{\Box}) = -\bar{\Box} + R_k(-\bar{\Box})$.
In this framework, we can easily verify that the Jacobian gives the following contributions to the flow equation
\begin{align}\label{rewritten Jacobian}
-\frac{1}{2} \Tr^\prime_{(1\text{T})} \Bigg[ \frac{\p_t P_k}{P_k -\bar R/4} \Bigg]
-\frac{1}{2} \Tr_{(0)}'' \Bigg[ \frac{\p_t P_k \left( 2\,P_k - \bar R /3 \right)}{ P_k\, \left(P_k- \bar R/3\right) } \Bigg]
+   \Tr_{(0)}' \bigg[ \frac{\p_t P_k}{P_k} \bigg] .
\end{align}

\subsection{Computing beta functions}
In this subsection, we present the basic strategy to calculate the beta functions. In the present work we use the gauge choice 
$\beta = 0$ and $\alpha \to 0$, for which a considerable simplification in the flow equations happens.
As discussed in Appendix~\ref{list of Hessian}, for this gauge choice, the $\sigma$ mode in the York decomposition \eqref{gravitondecomposition} is decoupled from the mixing with the other spin-0 scalar modes ($h$ and $\varphi$ - with $\varphi$ being the quantum fluctuation of the scalar field about a fixed background configuration).
It is important to emphasize, however, that beta functions are (off-shell) gauge dependent quantities, and therefore the physical conclusions should be carefully extracted from  these objects. Some investigation has been done in this direction from the FRG perspective for pure-gravity truncations in \cite{Gies:2015tca,deBrito:2018jxt} and gravity-matter approximation in \cite{Eichhorn:2016esv}. We mention that the analysis performed in \cite{Eichhorn:2016esv} indicates that the sign of critical exponent associated with the Yukawa coupling remains the same for different gauge choices.
One could expect this fact since the main contribution comes from the TT graviton mode which is independent from the gauge parameters.
In this paper we will not address the present question in more detail, however, we expect that, with respect to gauge dependence, our results do not present significant differences compared to \cite{Eichhorn:2016esv}.

For the system \eqref{general action} with the York decompositions \eqref{gravitondecomposition}, \eqref{ghostdecomposition} and this gauge choice, the Wetterich equation \eqref{Wetterich equation} reads
\begin{align}
\p_t \Gamma_k &= \frac{1}{2} \Tr_{(\text{2TT})} \left[ G_{\TT} \,\p_t \textbf{R}_k^{\TT}  \right] +
\frac{1}{2} \Tr_{(\text{0})} \left[ G_{\text{SS}} \, \p_t \textbf{R}_{k}^{\text{SS}}  \right] - \frac{1}{2} \Tr_{(1/2)} \left[ G_{\Psi \Psi} \, \p_t \textbf{R}_k^{\Psi \Psi} \right] + \Delta \mathcal{T}_{(1)} + \Delta \mathcal{T}_{(0)} + \nonumber\\[2ex]
&\quad
\,+\frac{1}{2}\Tr_{(0)} \left[ \Gamma^{(2)}_{\text{S} \Psi} \, G_{\Psi \Psi} \, \Gamma^{(2)}_{\Psi \text{S}} \, G_{\text{SS}} \,\p_t \textbf{R}_{k}^{\text{SS}} \,G_{\text{SS}} \right] 
-\frac{1}{2}\Tr_{(1/2)} \left[ \Gamma^{(2)}_{\Psi \text{S}} \, G_{\text{SS}} \,\Gamma^{(2)}_{\text{S} \Psi} \, G_{\Psi \Psi} \, \p_t \textbf{R}_{k}^{\Psi\Psi} \,G_{\Psi\Psi} \right] ,
\end{align}
where the first term on the right-hand side is the contribution from the TT graviton mode; the subscription ``SS" denotes contributions from $\varphi$ and $h$; the contributions from the fermion-loop corresponds to the third term; $\Delta \mathcal{T}_{(1)} $ and $\Delta \mathcal{T}_{(0)} $ are the spin-1 vectors ($\xi_\mu$, $C_\mu^\perp$ and the spin-1 contributions from the Jacobians) and the spin-0 scalars ($\sigma$, $C$ and the spin-0 contributions from the Jacobians), respectively; and the last two terms are contributions from mixings between bosonic and fermionic degrees of freedom.
The quantities $G_{\Phi_i \Phi_j}$ represent the regularized full propagator of the fields $\Phi_i$ and $\Phi_j$.
 For more details on the structure of the flow equation, see Appendix~\ref{structure of RG flow}. In what follows, it is convenient to rewrite the flow equation as
\begin{align} \label{decomposed Wetterich equation}
\p_t \Gamma_k =  \mathcal{T}_{(\text{2TT})} + \mathcal{T}_{(\text{1T})} + \mathcal{T}_{(0)}  + \mathcal{T}_{(1/2)} + \mathcal{T}_{\text{mixed}},
\end{align}
where the contributions $\mathcal{T}$'s are given by
\begin{align}
&\qquad\quad\mathcal{T}_{(\text{2TT})}=\frac{1}{2}\Tr_{(\text{2TT})} \left[ G_{\TT} \, \p_t \textbf{R}_k^{\TT}  \right]\,,\qquad
\mathcal{T}_{(1/2)}= - \frac{1}{2}\Tr \left[ G_{\Psi\Psi}\,\p_t \textbf{R}_{k}^{\Psi\Psi} \right]\,,\nn[2ex]
&\quad\quad\,\,\, \mathcal{T}_{(0)} =  \frac{1}{2} \Tr_{(0)} \left[ G_{\text{SS}} \, \p_t \textbf{R}_k^{\text{SS}}  \right] +\Delta \mathcal{T}_{(0)}\,,\qquad\qquad\quad
\mathcal{T}_{(\text{1T})}= \Delta \mathcal{T}_{(\text{1T})}  \,,
\nn[2ex]
&\mathcal{T}_{\text{mixed}}= \frac{1}{2}\Tr_{(0)}\left[ \Gamma^{(2)}_{\text{S} \Psi} \, G_{\Psi \Psi} \, \Gamma^{(2)}_{\Psi \text{S}} \, G_{\text{SS}} \,\p_t \textbf{R}_{k}^{\text{SS}} \,G_{\text{SS}}\right] 
- \frac{1}{2}\Tr_{(1/2)}\left[ \Gamma^{(2)}_{\Psi \text{S}} \, G_{\text{SS}} \, \Gamma^{(2)}_{\text{S} \Psi} \, G_{\Psi\Psi} \,\p_t \textbf{R}_{k}^{\Psi\Psi} \,G_{\Psi\Psi} \right] \,.
\end{align}
Their derivations and the explicit form of $\Delta \mathcal{T}_{(\text{1T})}$ and $\Delta \mathcal{T}_{(0)}$ are shown in Appendix~\ref{structure of RG flow}. 
Here, the subscripts on the trace stand for the spin of the corresponding fluctuation.
For explicit computations, the Hessians associated with the present truncation, which are necessary to compute the right-hand side of \eqref{decomposed Wetterich equation}, are listed in Appendix~\ref{list of Hessian}.
The results presented in this paper are computed in the background approximation, where we set the metric fluctuation to zero in the flow equation after computing the Hessians.
In this work we used Litim's optimized profile function (see Appendix \ref{list of Hessian}) \cite{Litim:2001up}.
We use the Type-I optimized cutoff ($z = -\bar \Box$) in order to compute $\mathcal{T}_{(2)}$, $\mathcal{T}_{(1)}$ and $\mathcal{T}_{(0)}$, while for the spin $1/2$, we employ the Type-II choice ($z = - \bar \Box + \bar R /4 \equiv - \Slash D^2$) in order to have the correct sign for the fermionic contributions to the Newton coupling constant~\cite{Dona:2012am}.
The computation of $\mathcal{T}_{\textmd{mixed}}$ is performed around flat background, in such a case  both the Type-I and -II cutoff choices collapses to $z = -\p^2$.

In order to extract beta functions from  the Wetterich equation, we can project \eqref{decomposed Wetterich equation} in different sectors of interest. For the gravitational couplings, the projection consists in setting all matter fields to zero. In such a case, we can easily verify that $\mathcal{T}_{\text{mixed}}$ does not give any contribution. Substituting our truncation in the left-hand side of  \eqref{decomposed Wetterich equation}, we find 
\begin{align}
&\frac{1}{16\pi G}  \int d^4 x\, \sqrt{\bar g} \left[ 2 \left(\p_t \Lambda + \eta_G \, \Lambda \right) - \eta_G\,\bar{R} + ( \eta_G \, a  + \p_t a ) \,\bar{R}^2 \,+ \right. \nonumber \\
&\left. \qquad\qquad\qquad +\,  ( \eta_G \, b  + \p_tb ) \,\bar R_{\mu\nu} \bar R^{\mu\nu} \right]
= \left[ \mathcal{T}_{(\text{2TT})} + \mathcal{T}_{(\text{1T})} + \mathcal{T}_{(0)}  + \mathcal{T}_{(1/2)} \right]_{\text{Proj. Gravity} } ,
\end{align}
where $\eta_G = - \, G^{-1} \p_t G$. In terms of the dimensionless quantities, the last equation can be rewritten in the following way
\begin{align}\label{Flow_Grav}
&\frac{k^4}{16\pi \tilde{G}}  \int d^4 x\, \sqrt{\bar g} \left[ 2 \tilde{\Lambda}\, \left( \tilde{\Lambda}^{-1} \beta_{{\Lambda}} - \tilde{G}^{-1} \beta_{{G}} + 4  \right) + \left( \tilde{G}^{-1} \beta_{{G}} - 2 \right) \bar{R} + 
\left(  \beta_{{a}} - \tilde{G}^{-1} \, \tilde{a}\, \beta_{{G}}  \right) \bar{R}^2 \,+\right. \nonumber \\
&\left. \qquad\qquad\qquad\qquad+  
\left(  \beta_{{b}} - \tilde{G}^{-1} \, \tilde{b}\, \beta_{{G}}  \right) \bar R_{\mu\nu} \bar R^{\mu\nu} \right]
= \left[ \mathcal{T}_{(\text{2TT})} + \mathcal{T}_{(\text{1T})} + \mathcal{T}_{(0)}  + \mathcal{T}_{(1/2)} \right]_{\text{Proj. Gravity} } ,
\end{align}
where $\tilde{\Lambda} = k^{-2} \Lambda$, $\tilde{G} = k^{2} G$, $\tilde{a} = k^{2} a$ and $\tilde{b} = k^{2} b$ are dimensionless couplings. In addition,  we define the beta function of the dimensionless couplings as $\beta_{\Lambda} = \p_t \tilde{\Lambda}$, $\beta_{G} = \p_t \tilde{G}$, $\beta_{a} = \p_t \tilde{a}$ and $\beta_{b} = \p_t \tilde{b}$. By expanding the left-hand side of the above equation eq.\,\eqref{Flow_Grav} in powers of curvature invariants, we can compute beta functions by matching the coefficients of the same invariants.

In order to project the flow equation to the scalar field potential, we set the curvature terms and the fermionic fields to zero.
In this case, we only have contributions coming from $\mathcal{T}_{(\text{2TT})}$, $\mathcal{T}_{(0)}$ and $\mathcal{T}_{(1/2)}$. 
Then, the flow equation takes the form
\begin{align}
\int d^4 x\, \sqrt{\bar g}\, \p_t V(\phi^2) = 
 \left[ \mathcal{T}_{(\text{2TT})}  + \mathcal{T}_{(0)}  + \mathcal{T}_{(1/2)} \right]_{\text{Proj. Scalar}} .
\end{align}
Defining the renormalized field $\tilde{\phi} = Z_\phi^{1/2} \phi$, where $Z_\phi$ denotes the wave function renormalization, the flow equation for the scalar potential can be recast in the following way
\begin{align}\label{flow_renormalized_potential}
\int d^4 x\, \sqrt{\bar g}\, \left( \p_t\tilde{V}(\tilde{\phi}^2) +4\tilde V(\tilde \phi) - \eta_\phi \, \tilde{\phi}^2 \,\tilde{V}'(\tilde{\phi}^2) \right) = 
k^{-4}\left[ \mathcal{T}_{(\text{2TT})}  + \mathcal{T}_{(0)} + \mathcal{T}_{(1/2)} \right]_{\text{Proj. Scalar} \,= \,0} ,
\end{align}
where $\tilde{V}(\tilde{\phi}^2) = V(\phi^2)/k^4$ denotes the dimensionless scalar potential. Expanding the dimensionful scalar potential as $V(\phi^2) = \sum_{n=1}^\infty \lambda_{2n} \,\phi^{2n}$, the corresponding dimensionless potential is given by $\tilde{V}(\tilde{\phi}^2) = \sum_{n=1}^\infty k^{4-2n} \, \tilde{\lambda}_{2n} \tilde{\phi}^{2n}$, where we have defined the dimensionless couplings as $\tilde{\lambda}_{2n} = k^{2n-4} Z_\phi^{-n} \lambda_{2n}$ and the renormalized scalar field is $\tilde \phi = Z_\phi^{1/2} \phi$. By expanding both sides of \eqref{flow_renormalized_potential} as a power series in $\tilde{\phi}^2$, we can extract beta functions for the dimensionless coupling constants.

Finally, in order to compute beta functions for the Majorana masses and the Yukawa coupling, we set the curvature terms to zero and the matter fields to constant configurations. Also, we discard all contributions of order $\mathcal{O}(\tilde{\phi}^2)$ or higher. Implementing such projections, the beta function associated with the Majorana-right (left) mass can be extracted by keeping only those terms proportional to $ {\bar \psi}^c P_{R(L)} \psi  + \textmd{h.c.} $, while the running of the Yukawa coupling can read from those terms proportional to $\phi \, \bar \psi \psi$. 
We have
\begin{align}
\frac{1}{2}\int d^4 x\, \sqrt{\bar g} \left[ \p_t m_M^{R(L)} ( {\bar \psi}^c P_{R(L)} \psi  + \textmd{h.c.} )\right]
= \left[ \mathcal{T}_{(\text{2TT})} +  \mathcal{T}_{(0)}  + \mathcal{T}_{ \text{mixed} } \right]_{\text{Proj. Maj. Right (Left)} } ,
\end{align}
and
\begin{align}
\int d^4 x\, \sqrt{\bar g} \,\, \p_t y_D \, \phi \, {\bar \psi}  \psi 
= \left[ \mathcal{T}_{(\text{2TT})} +  \mathcal{T}_{(0)}  + \mathcal{T}_{ \text{mixed} } \right]_{\text{Proj. Yukawa} } .
\end{align}
In terms of renormalized fields $\tilde \phi = Z_\phi^{1/2} \phi$, $\tilde{\psi} = Z_\psi^{1/2} \psi$ and $\tilde{\bar \psi} = Z_\psi^{1/2} \bar \psi$, we define dimensionless couplings $\tilde{y}_D = Z_\phi^{-1/2} Z_\psi^{-1}\, y_D$ and $\tilde{m}_{M}^{R(L)} = k^{-1} Z_\psi^{-1} m_{M}^{R(L)}$ so that 
\begin{align}\label{Flow_Majorana}
\left( \beta_{m_M^{R(L)}} \!+\! (1- \eta_\psi) \tilde{m}_M^{R(L)} \right)\!\int d^4 x\, \sqrt{\bar g} \left[ {\tilde{\bar \psi}}^{\,c} P_{R(L)} \tilde{\psi}  + \textmd{h.c.} \right]
= 2\,k^{-1}  \left[ \mathcal{T}_{(\text{2TT})} +  \mathcal{T}_{(0)}  + \mathcal{T}_{ \text{mixed} } \right]_{\text{Proj. Maj. Right (Left)} } ,
\end{align}
and
\begin{align}\label{Flow_Yukawa}
\left[ \beta_y - \left( \eta_\psi+\frac{1}{2}\eta_\phi\right) \tilde{y}_D \right]  \int d^4 x\, \sqrt{\bar g} \, \tilde{\phi} \, \tilde{\bar \psi}  \tilde{\psi} 
= \left[ \mathcal{T}_{(\text{2TT})} +  \mathcal{T}_{(0)}  + \mathcal{T}_{ \text{mixed} } \right]_{\text{Proj. Yukawa} } ,
\end{align}
where we have defined the beta functions $\beta_{m_M^{R}} = \p_t \tilde{m}_M^{R}$, $\beta_{m_M^{L}} = \p_t \tilde{m}_M^{L}$ and $\beta_y = \tilde{y}_D$. The r.h.s. of the \eqref{Flow_Majorana} and \eqref{Flow_Yukawa} can be organized in terms of diagrams as one can see in Figs. \ref{Feynman diagrams of Majorana beta functions} and \ref{Feynman diagrams of Yukawa beta functions}.

\section{Results}\label{result section}

\subsection{Contributions from graviton fluctuations on the Majorana masses \label{results_1}}

Let us start from the discussion on the impact of quantum gravity fluctuations on the running of the Majorana masses. From the results reported in the Appendix \ref{Explicit form of beta functions} we observe that the beta function of the left-handed Majorana mass has the following structure
\begin{align}
\beta_{m_M^{L}} = (-1+\eta_\psi)\tilde m_M^L + \big(\,  {\mathcal A_{1,L}} {\tilde m}_M^L+{\mathcal A_{3,L}}  (\tilde m_M^L)^3\, \big)\tilde G
+{\mathcal M}_{y^2,L}\,  \tilde y_D^2\tilde m_M^R,
\label{schematic beta function of majorana mass}
\end{align}
where $\mathcal A_{1,L}$, $\mathcal A_{3,L}$ and ${\mathcal M}_{y^2,L}$ are computed in terms of threshold functions (see Appendix \ref{Explicit form of beta functions}). The beta function for the right-handed Majorana mass can be obtained by exchanging $L \leftrightarrow R$ in the above expression. For the sake of simplicity, we will not take into account effects coming from the fermionic anomalous dimension in our discussion.

From the perspective of asymptotically safe quantum gravity our goal is to understand whether the running of the Majorana masses is compatible with an UV complete setting. In this framework, we have some interesting possibilities:
\begin{itemize}
	\item[i)] UV fixed point with $\tilde{m}_M^R = \tilde{m}_M^L = 0$;
	\item[ii)] UV fixed point with $\tilde{m}_M^R =0$ and $\tilde{m}_M^L \neq 0$ (or $\tilde{m}_M^R \neq 0$ and $\tilde{m}_M^L = 0$);
	\item[iii)] UV fixed point with $\tilde{m}_M^R \neq 0$ and $\tilde{m}_M^L \neq 0$.
\end{itemize}

Let us focus on the first scenario. In this case it is straightforward to see that $\tilde{m}_M^R = \tilde{m}_M^L = 0$ is a fixed point of the subsystem $\beta_{m_M^{L}} = 0$ and $\beta_{m_M^{R}} = 0$, irrespective of the values of the gravitational couplings. In the absence of gravity the critical exponents associated with this fixed point correspond to the canonical scaling dimension of the Majorana masses and, therefore, are positive. If we take quantum gravity fluctuations into account, the critical exponents associated with Majorana masses are given by
\begin{align}
\theta^{M}_1 = 1 -  \mathcal A_{1} \tilde{G}^* - \frac{1}{8\pi^2}  \tilde y^{*2}_D \qquad
\textmd{and}\qquad
\theta^{M}_2 = 1 -  \mathcal A_{1} \tilde{G}^* + \frac{1}{8\pi^2}  \tilde y^{*2}_D ,
\end{align} 
where $\mathcal A_{1} = \mathcal A_{1,L}|_{\tilde{m}_M^L = 0} = \mathcal A_{1,R}|_{\tilde{m}_M^R = 0}$ and we have used ${\mathcal M}_{y^2,L}|_{\tilde{m}_M^R = 0} = {\mathcal M}_{y^2,R}|_{\tilde{m}_M^L = 0} = (8\pi^2)^{-1}$. Moreover, $\tilde{G}^*$ and $\tilde y^{*}_D$ correspond, respectively, to the values of the dimensionless Newton's constant and the Yukawa coupling at the fixed point. The relevant question here is whether quantum gravity contributions could change the sign of the critical exponents, driving (one of) them towards irrelevance. In order to answer this question we first note that, at the fixed point with $\tilde{m}_M^R = \tilde{m}_M^L = 0$, we can express $\mathcal A_{1} = \mathcal{D}_y^*$ (see Appendix \ref{Explicit form of beta functions} for the definition of $\mathcal{D}_y$). As we shall see in the next section, the viability of UV completion in the Yukawa sector requires $\mathcal{D}_y^* < 0$. We first explore a scenario with $\tilde y^{*}_D = 0$. In this case the critical exponents for the Majorana masses become
\begin{align}
\qquad \theta^{M}_{1} = \theta^{M}_{2} = 1 -  \mathcal D_{y}^* \, \tilde{G}^* \,, \qquad\qquad\qquad  (\textmd{with} \,\,\tilde y^{*}_D = 0).
\end{align}
In the case of a fixed point with non-vanishing Yukawa coupling we have $y^{*}_D = \sqrt{ - 8\pi^2\, \tilde G^{*}\, \tilde{\mathcal{D}}_{y}^* }$ (see section \ref{results_2}) and, as a consequence, we find the following critical exponents
\begin{align}
\theta^{M}_1 = 1 \qquad
\textmd{and}\qquad
\theta^{M}_2 = 1 - 2\,\mathcal D_{y}^* \, \tilde{G}^*  \qquad\qquad  (\textmd{with} \,\,\tilde y^{*}_D \neq 0) .
\end{align} 
Assuming $\tilde{G}^* > 0$, the requirement $\mathcal D_{y}^* < 0$, which is necessary for UV completion in the Yukawa sector, restricts the space of viable fixed point values in such a way that the critical exponents for the Majorana masses (at the fixed point value characterized $\tilde{m}_M^R = \tilde{m}_M^L = 0$) remains positive after the inclusion of quantum gravity effects. This is the simplest (non-trivial) UV complete extension of a Higgs-Yukawa model incorporating Majorana masses. If $\theta^{M}_{1,2} < 0$ due to quantum gravity fluctuations, the flow of the Majorana masses would be trivial. In summary, for the possibility (i), demanding a UV-complete Yukawa sector necessarily implies that the Majorana masses will be relevant parameters even at non-trivial values for the other couplings fixed-point values.

The analysis regarding the possibility of the fixed points (ii) and (iii) is more complicated. In this case the fixed point equations are more involved due to non-polynomial dependence on the Majorana masses and, as consequence, there is no simple criteria to probe the existence of such fixed points analytically. In the Appendix \ref{Fixed point analysis} we have shown the possibility of finding numerical solutions corresponding to fixed points in these classes. Instead of looking for further numerical solutions, in the next section we assume the existence of fixed point solutions featuring $\tilde{m}_M^R \neq 0$ and/or $\tilde{m}_M^L \neq 0$ and investigate their consequences to the viability of UV completion for the Higgs-Yukawa sector as well as in the predictability of the Higgs mass in the framework of asymptotically safe quantum gravity.

\subsection{Impact of Majorana masses on the Higgs-Yukawa couplings \label{results_2}}
A fruitful strategy to  probe quantum gravity induced effects in the UV regime of the matter sector is to treat the fixed points values of the gravitational couplings as free parameters in the beta functions associated with matter couplings. This approach allows us to explore the viability of quantum gravity induced UV completion and predictive RG trajectories for the matter sector~\cite{Eichhorn:2017eht}. Although the computation of beta functions associated with the matter sector involves a given truncation for the effective average action, this strategy gives information beyond  a specific choice of truncation, since the values for the gravitational coupling at the fixed points are not restricted by the chosen approximation.
Nevertheless, in Appendix~\ref{Fixed point analysis}, we show results for fixed point values and critical exponents in a maximally symmetric background.
In the present section, we adopt the same reasoning in order to investigate the impact of Majorana masses on Yukawa-Higgs systems in the UV regime. 
Moreover, we treat the fixed point values of the Majorana masses as free parameters and analyze their impact  to the beta functions of the Yukawa and $\lambda_4 \phi^4$ interactions.

We start  by discussing the fixed point scenarios for the Yukawa sector. As discussed in the recent literature, quantum gravity fluctuations play an important role in the running of the Yukawa coupling \cite{Zanusso:2009bs,Oda:2015sma,Eichhorn:2016esv,Christiansen:2017qca,Eichhorn:2017eht,Hamada:2017rvn}, leading to an interesting phenomenological scenario with the prediction of the top quark mass~\cite{Eichhorn:2017ylw}, as well as, top-bottom mass difference~\cite{Eichhorn:2018whv}. Earlier works on higher-derivative gravity also indicates quantum gravity induced UV completion for the Yukawa coupling \cite{Shapiro:1989dq,Buchbinder:1989jd,Buchbinder:1992rb}. 

The beta function of the Yukawa coupling has the following structure,
\begin{align}
\beta_y=
\left(\frac{\eta_\phi}{2}+ \eta_\psi \right) \tilde y_D + \left( \mathcal{D}_y + \mathcal{D}_{y,\lambda_2} \, {\tilde\lambda}_2   
+  \mathcal{D}_{y,M}^{(1)}\, ( {\tilde m}_M^L+{\tilde m}_M^R )^2   
+  \mathcal{D}_{y,M}^{(2)}\, {\tilde m}_M^R {\tilde m}_M^L  \right)\tilde G \, \tilde y_D + \mathcal{M}_{y^3} \,{\tilde y_D}^3 \, ,
\end{align}
where we define the $\mathcal{D}_{y}$'s and $\mathcal{M}_{y^3}$ in terms of threshold functions (see Appendix \ref{Explicit form of beta functions}). In order to simplify our analysis, we consider some additional approximations: i) we neglect contributions coming from the anomalous dimension; ii) we assume the Gaussian fixed point value for $\tilde{\lambda}_2$. Taking these approximations into account, the beta function for the Yukawa coupling constant reads
\begin{align}
\beta_y= \tilde{\mathcal{D}}_{y} \,  \tilde G \, \tilde y_D + \mathcal{M}_{y^3} \,{\tilde y_D}^3 ,
\end{align}
where $\tilde{\mathcal{D}}_{y} = \mathcal{D}_y  + \mathcal{D}_{y,M}^{(1)} \, ( {\tilde m}_M^L+{\tilde m}_M^R )^2   
+  \mathcal{D}_{y,M}^{(2)}\, {\tilde m}_M^R {\tilde m}_M^L  $.
Following the discussion presented in reference~\cite{Eichhorn:2017eht}, there are two possibilities for quantum gravity induced UV completion in the Yukawa sector. The first one is the Yukawa-free fixed point,  i.e., $\tilde{y}_D^* = 0$. In this case, the critical exponent associated with the Yukawa coupling constant is given by
\begin{align}
\theta_y \simeq - \left( \frac{\p \beta_y}{\p \tilde{y}_D} \right)_{\!*} = - \tilde{\mathcal{D}}_{y}^* \, \tilde G^{\,*} \, ,
\end{align}
where the notation $(\cdots)_{*}$ indicates that the stability matrix is evaluated at the fixed point. Assuming $\tilde{G}^{\,*} > 0$, we conclude that the Yukawa coupling becomes a relevant direction ($\theta_y > 0$) provided that\footnote{The possibility $\tilde{y}_D^* = 0$ and $\theta_y < 0$ trivializes the Yukawa flow and, therefore, is not interesting from the physical point of view.} $\tilde{\mathcal{D}}_{y}^*  < 0 $.
This is the case of asymptotically free UV completion for the Yukawa sector. 

The second scenario for Yukawa UV completion is characterized by an interacting fixed point induced by quantum gravity contributions, namely
\begin{align}
\tilde{y}_D^* = \sqrt{ - {\tilde G^{*}\, \tilde{\mathcal{D}}_{y}^*\,}/{\mathcal{M}_{y^3}^* } \,} \, .
\label{non-trivial fixed point of Yukawa coupling constant}
\end{align}
The existence of such a fixed point is subject, of course, to the positivity of the square root argument. For small values of the Majorana masses at the fixed point, as we shall see later, it is possible to verify that $\mathcal{M}_{y^3}^* > 0$ and, therefore, the condition for existence of this fixed point is given by $\tilde{\mathcal{D}}_{y}^*  < 0 $.  More generally, the viability condition for such a scenario is given by $\tilde{\mathcal{D}}_{y}^* / \mathcal{M}_{y^3}^* < 0$. The critical exponent associated with this fixed point reads 
\begin{align}
\theta_y \simeq - \left( \frac{\p \beta_y}{\p \tilde{y}_D} \right)_{\!*} = 2 \,\tilde{\mathcal{D}}^{\ast}_{y} \,\tilde{G}^{\ast}\, .
\end{align}
If $\mathcal{M}_{y^3}^* > 0$, the existence of the fixed point \eqref{non-trivial fixed point of Yukawa coupling constant} is subject to $\tilde{\mathcal{D}}_{y}^*  < 0 $, implying $\theta_y < 0$ (UV repulsive fixed point). Such a scenario is particularly interesting since the requirement of UV completion allows us to make predictions in the IR regime~\cite{Eichhorn:2017eht}. On the other hand, if $\mathcal{M}_{y^3}^* < 0$, which can achieved for large values of the Majorana masses at the fixed point, the viability condition requires  $\tilde{\mathcal{D}}_{y}^* > 0$, resulting in $\theta_y > 0$ (UV attractive fixed point), so that the Yukawa coupling remains a free parameter.

\begin{figure}
	\begin{center}
		\includegraphics[width=7cm]{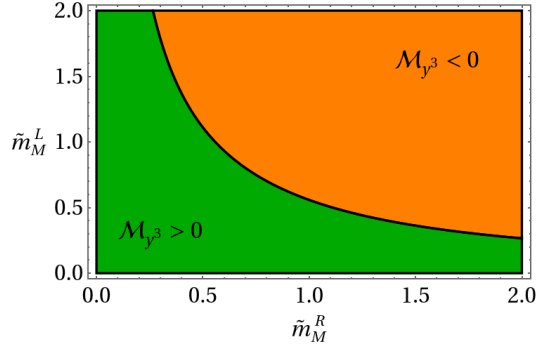}$\quad\qquad\qquad$
		\caption{Sign of $\mathcal M_{y^3}$ as a function of the Majorana masses. The green (darker) region correspond to $\mathcal M_{y^3} > 0$, while the orange (brighter) region indicates $\mathcal M_{y^3} < 0$.}
		\label{Plot-M_y^3}
	\end{center}
\end{figure}

It is interesting to analyze the behavior of $\tilde{\mathcal{D}}_{y}$ and $\mathcal{M}_{y^3}$ as a function of the Majorana masses and the gravitational parameters. For the sake of simplicity, let us start with the TT-approximation, namely, we take only the TT graviton effects into account.
In this case we have 
\begin{align}
	\tilde{\mathcal{D}}_{y} = \frac{5\,(2+3\tilde{b})}{4\pi \,(1+\tilde{b}-2 \tilde\Lambda)^2} .
\end{align}
The $\mathcal{M}_{y^3}$ term, which only receives contributions from matter fluctuations, is given by
\begin{align}
	\mathcal{M}_{y^3} = \frac{2 + (\tilde{m}_M^R - \tilde{m}_M^L)^2  
		- \tilde{m}_M^R\,\tilde{m}_M^L \left( 1 + 2 \, (\tilde{m}_M^R)^2 + 2\,(\tilde{m}_M^L)^2 + (\tilde{m}_M^R)^2 (\tilde{m}_M^L)^2\right)}{16\pi^2 (1 + (\tilde{m}_M^R)^2) (1 + (\tilde{m}_M^L)^2) } .
\end{align}
Within the TT-approximation, the sign of $\tilde{\mathcal{D}}_{y}$ is completely determined by the value of the gravitational coupling $\tilde{b}$, namely, $\tilde{\mathcal{D}}_{y} < 0$ for $\tilde{b} < -2/3$ and $\tilde{\mathcal{D}}_{y} > 0$ for $\tilde{b} > -2/3$. The sign of $\mathcal{M}_{y^3}$ as a functions of the Majorana masses is depicted in Fig.\,\ref{Plot-M_y^3}.
It can be roughly approximated by $\mathcal{M}_{y^3} > 0$ for $\tilde{m}_M^R \tilde{m}_M^L \lesssim 1/2$ and $\mathcal{M}_{y^3} < 0$ for $\tilde{m}_M^R \tilde{m}_M^L \gtrsim 1/2$. It is interesting to observe that if we set one of the Majorana masses to zero, $\mathcal{M}_{y^3}$  remains positive even for arbitrary large values of the other Majorana mass.

\begin{figure}[htb!]
	\begin{center}
		\includegraphics[width=7cm]{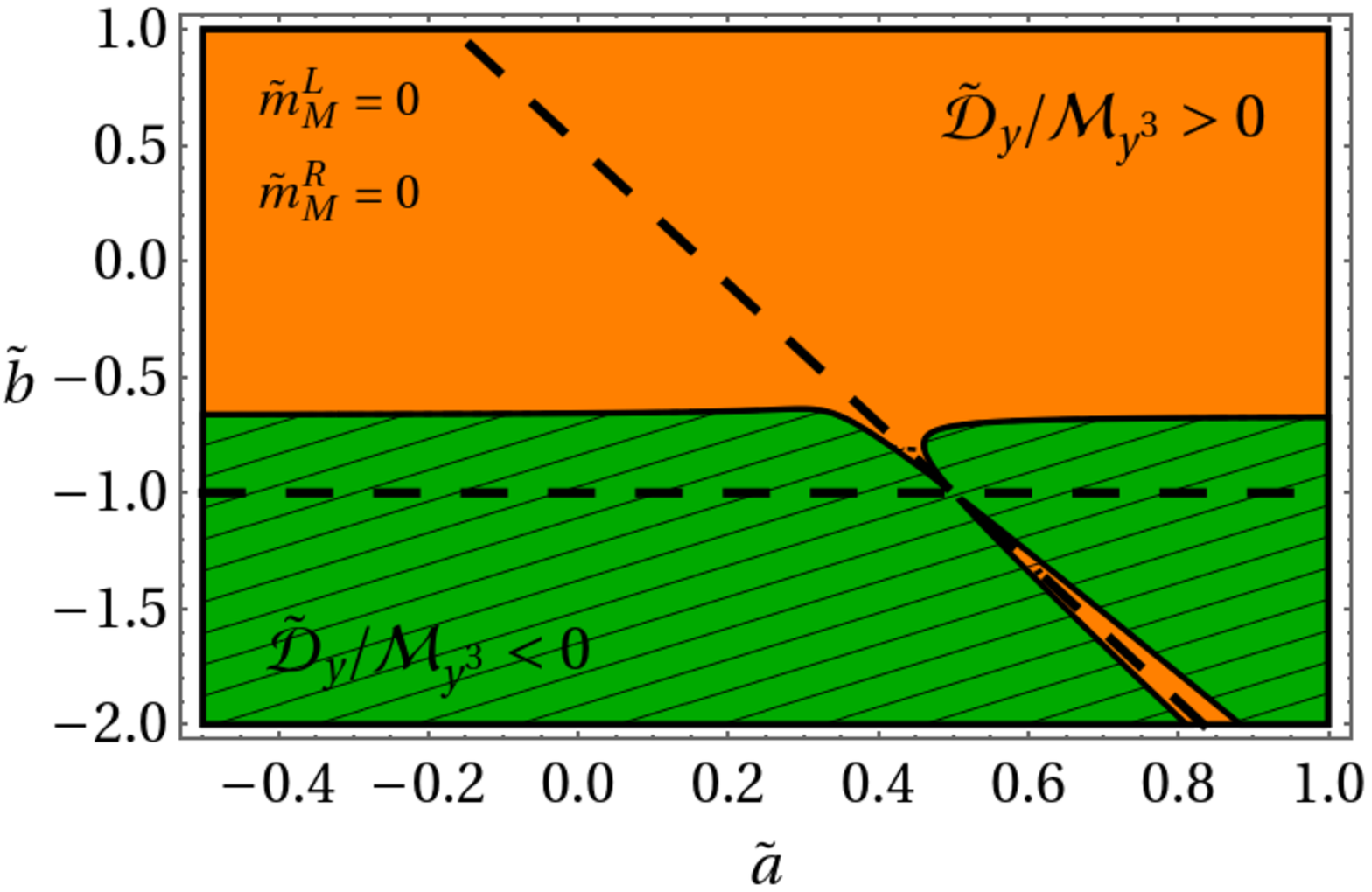} \qquad \includegraphics[width=7cm]{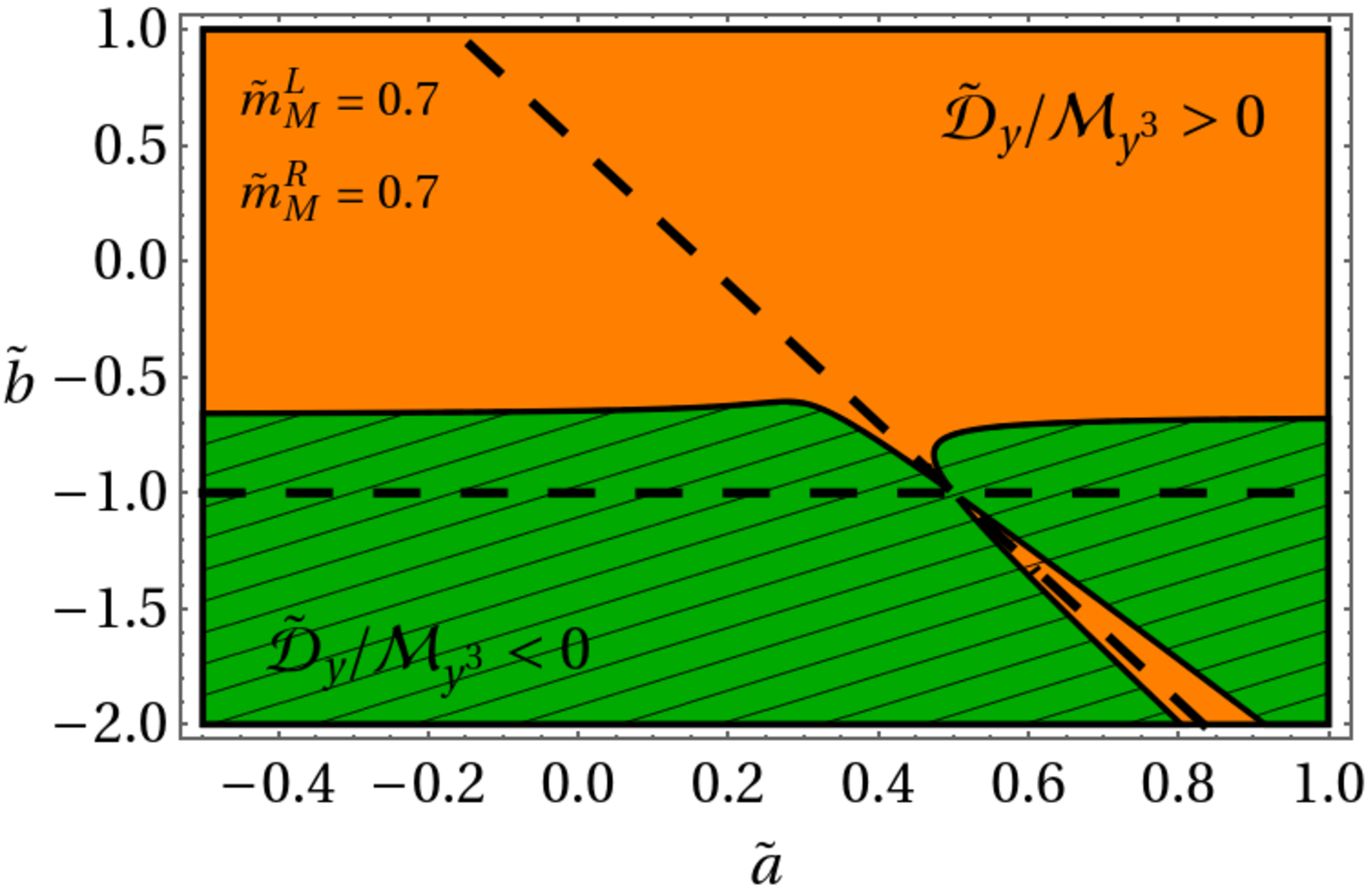} $\qquad\quad$ \\
		\includegraphics[width=7cm]{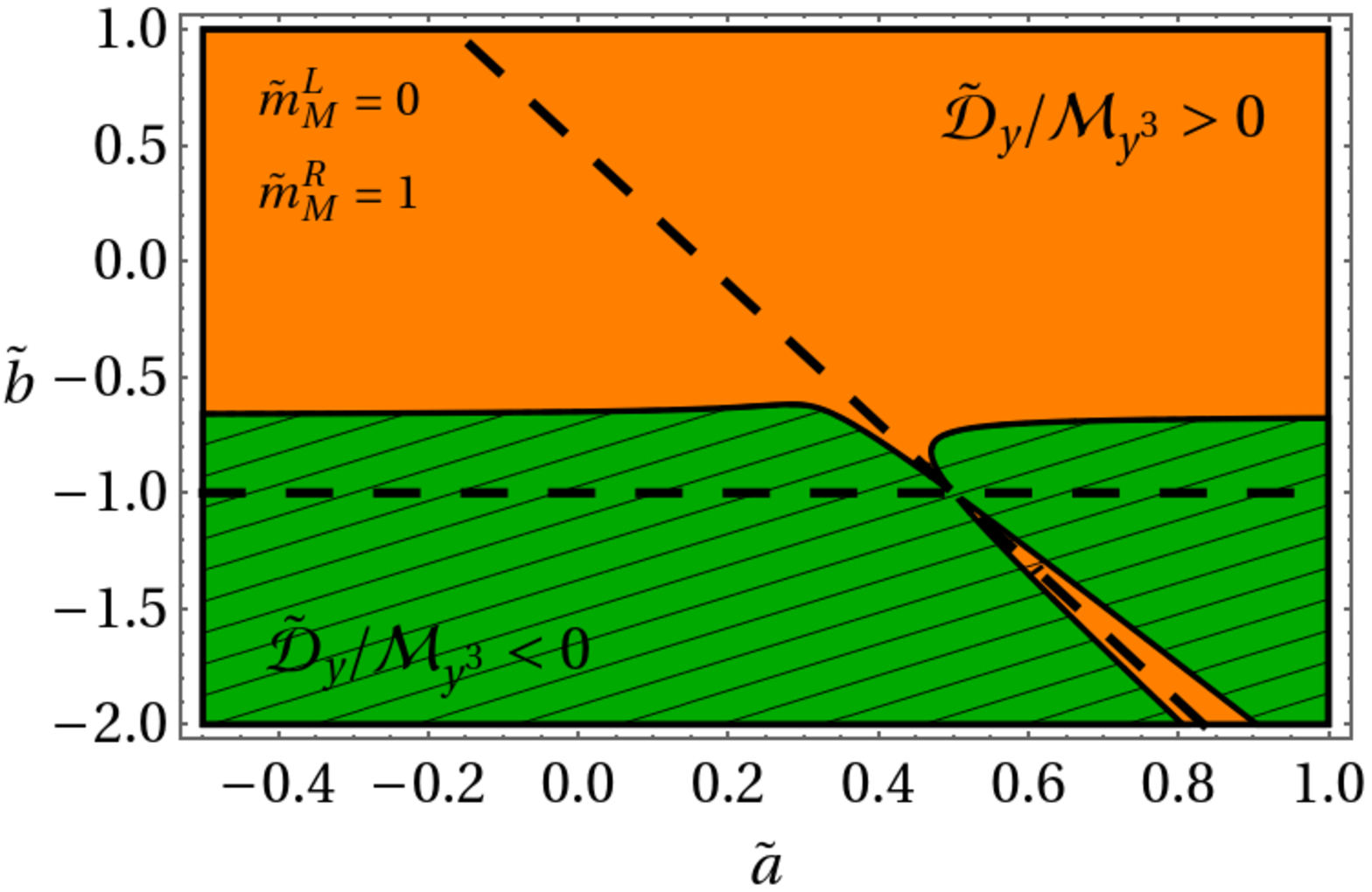} \qquad \includegraphics[width=7cm]{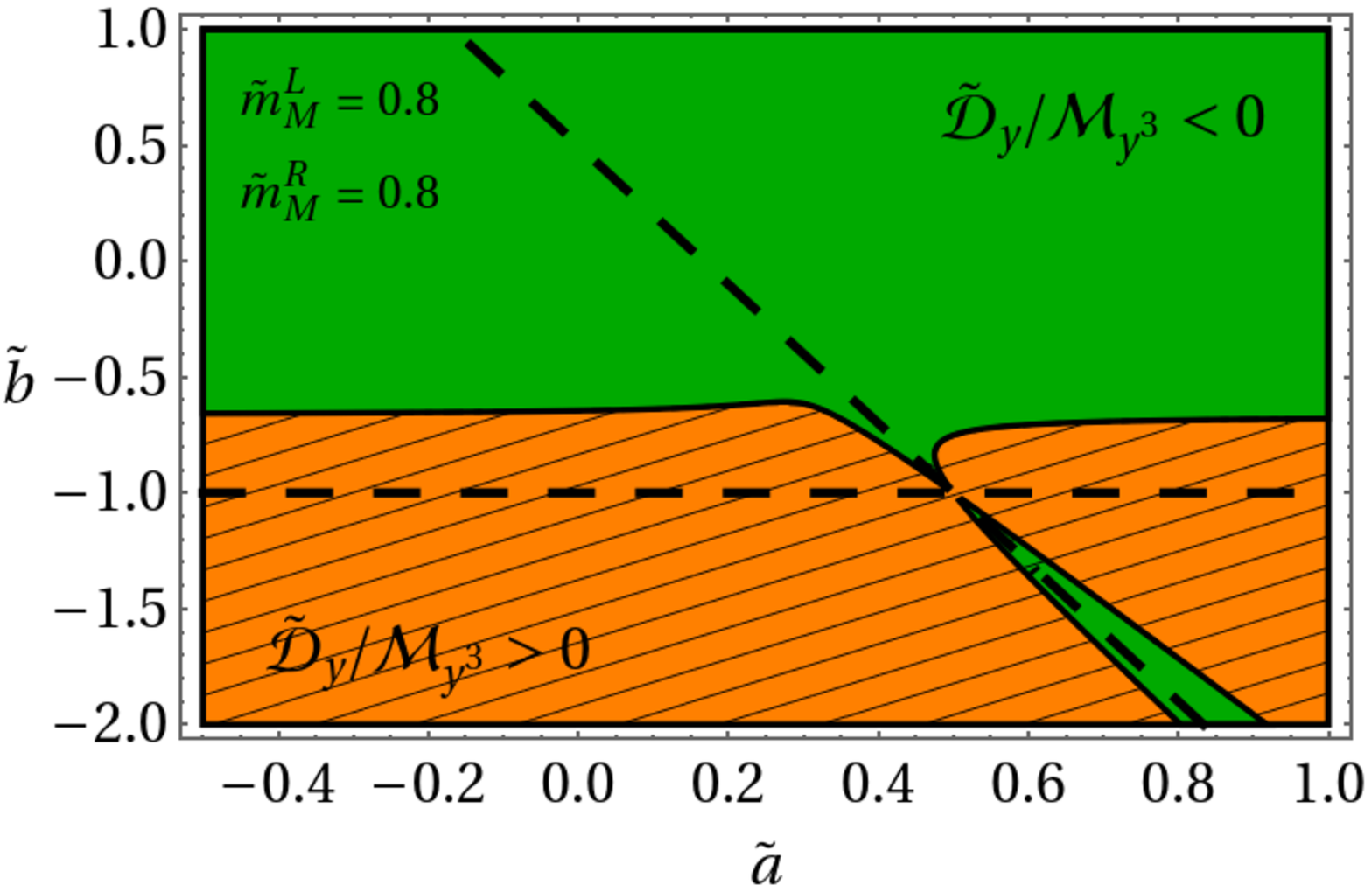} $\qquad\quad$
		\caption{Region plots indicating the sign $\mathcal{\tilde{D}}_y/\mathcal{M}_{y^3}$ as function of of the curvature squared couplings ($\tilde a$ and $\tilde b$). In all cases we consider the dimensionless cosmological constant as being zero. Different plots correspond to different choices of the Majorana masses. The green (darker) region correspond to $\mathcal{\tilde{D}}_y/\mathcal{M}_{y^3} < 0$, while the orange (brighter) region indicates $\mathcal{\tilde{D}}_y/\mathcal{M}_{y^3}>$. The mesh lines stand for the region constrained only by ${\tilde{\mathcal D}}_y < 0$. The dashed lines indicate the poles $1 +\tilde b = 0$ and $1 -6\tilde a -2\tilde b = 0$.
		}
		\label{Viable_Yukawa_a_x_b_several_masses}
	\end{center}
\end{figure}

Beyond the TT-approximation the situation is more complicated, since $\tilde{\mathcal{D}}_y$ depends on the parameters $\tilde{a}$, $\tilde{b}$, $\tilde{\Lambda}$, $\tilde{m}_M^R$ and $\tilde{m}_M^L$ (see Appendix~\ref{Explicit form of beta functions} for explicit expressions). We start investigating its behavior  by looking at the $(\tilde{a},\tilde{b})$-plane. In the panel exhibited in Fig.\,\ref{Viable_Yukawa_a_x_b_several_masses}, we show the sign of $\mathcal{\tilde{D}}_y/\mathcal{M}_{y^3}$ as a function of $\tilde{a}$ and $\tilde{b}$, for some fixed values of the Majorana masses. As one can observe, as long as the Majorana masses respect the inequality $\tilde{m}_M^R \tilde{m}_M^L \lesssim 1/2$ the viable region for the UV completion of the Yukawa sector remains qualitatively unchanged. On the other hand, for $\tilde{m}_M^R \tilde{m}_M^L \gtrsim 1/2$, there is a significant qualitative change of behavior, produced by the flip of sign in $\mathcal{M}_{y^3}$. It is important to emphasize that in all cases the sign of $\tilde{\mathcal{D}}_y$ does not feature important qualitative changes due to variations of the Majorana masses.

\begin{figure}
	\begin{center}
		\includegraphics[width=7cm]{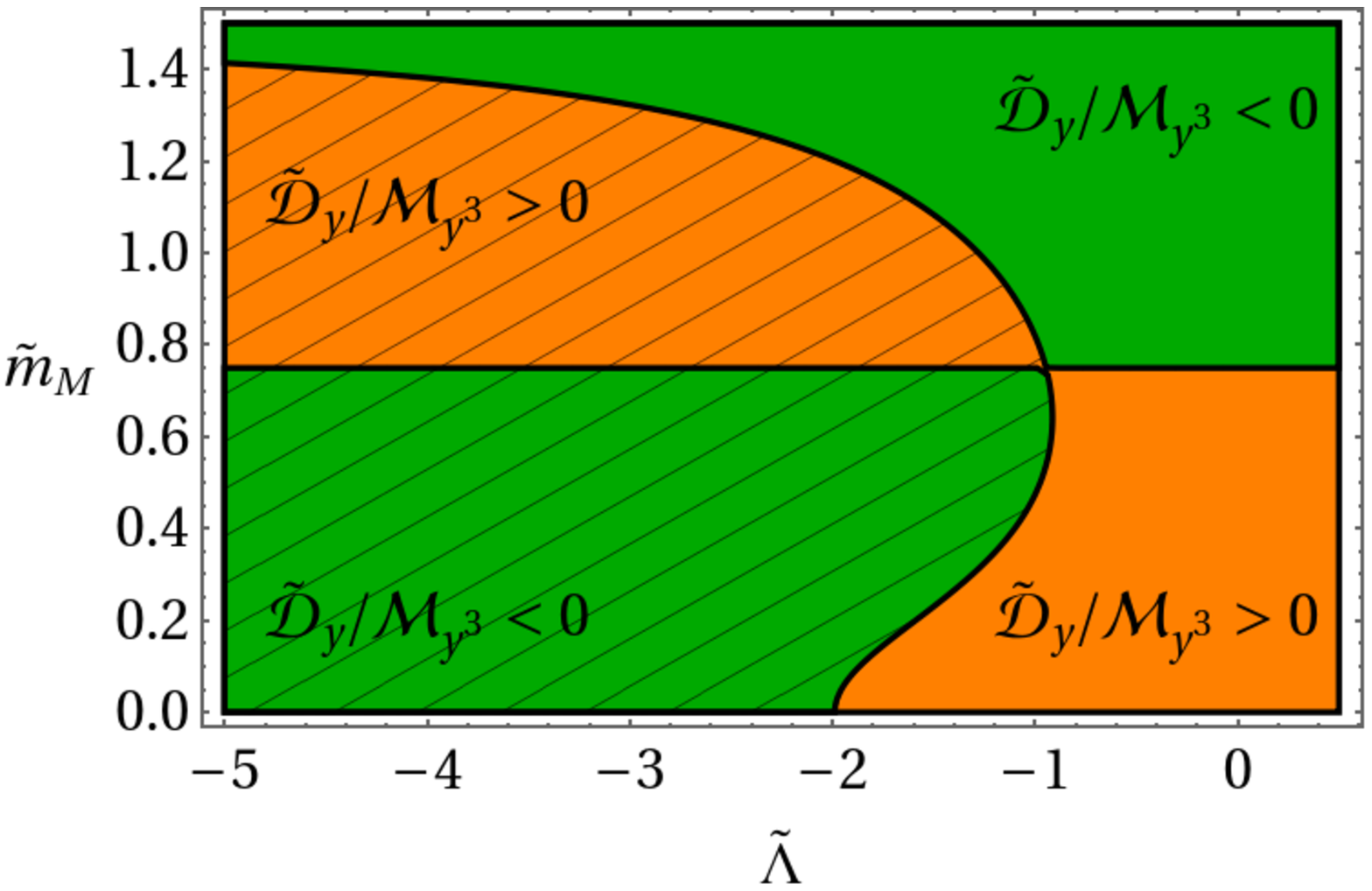} \qquad \includegraphics[width=6.75cm]{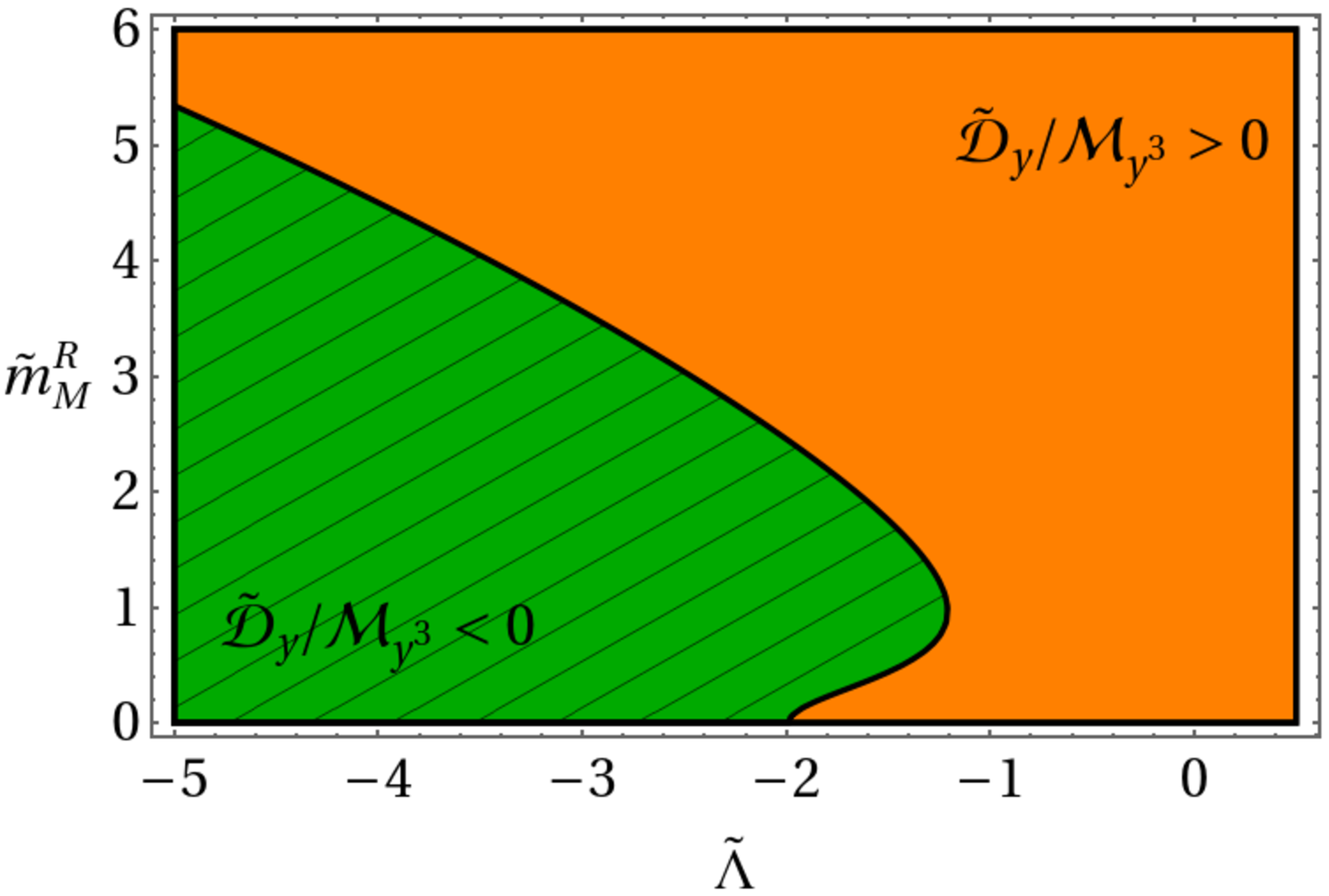}$\qquad\qquad$
		\caption{Region plots showing the sign of $\mathcal{\tilde{D}}_y/\mathcal{M}_{y^3}$ (with $\tilde a=\tilde b=0$) in terms of $\tilde \Lambda$ and the Majorana masses. The plot on the left we consider both Majorana masses as being equal ($\tilde{m}_M := \tilde{m}_M^R = \tilde{m}_M^L$) and the plot on the right correspond to the situation where one of the masses is zero ($\tilde{m}_M^L=0$). The green (darker) region correspond to $\mathcal{\tilde{D}}_y/\mathcal{M}_{y^3} < 0$, while the orange (brighter) region indicates $\mathcal{\tilde{D}}_y/\mathcal{M}_{y^3}>0$. The mesh lines stand for the region constrained only by ${\tilde{\mathcal D}}_y < 0$.}
		\label{Viable_Yukawa_Lambda_x_m}
	\end{center}
\end{figure}

Those plots represented in Fig.\,\ref{Viable_Yukawa_a_x_b_several_masses} are generated by setting the (dimensionless) cosmological constant to zero. It is interesting to investigate the case of non-vanishing cosmological constant combined with effects of Majorana masses. 
The left panel of Fig.\,\ref{Viable_Yukawa_Lambda_x_m} shows the sign of $\mathcal{\tilde{D}}_y/\mathcal{M}_{y^3}$ in the plane $(\tilde{\Lambda},\tilde{m}_M)$, where $\tilde{m}_M := \tilde{m}_M^R = \tilde{m}_M^L$, while in the right-hand side panel of Fig.\,\ref{Viable_Yukawa_Lambda_x_m} we show the case where one of the Majorana masses is set to zero (namely, $\tilde{m}_{M}^L = 0$). 
For both cases we set $\tilde{a} = \tilde{b} = 0$.

\begin{figure}[htb!]
	\begin{center}
		\includegraphics[width=7cm]{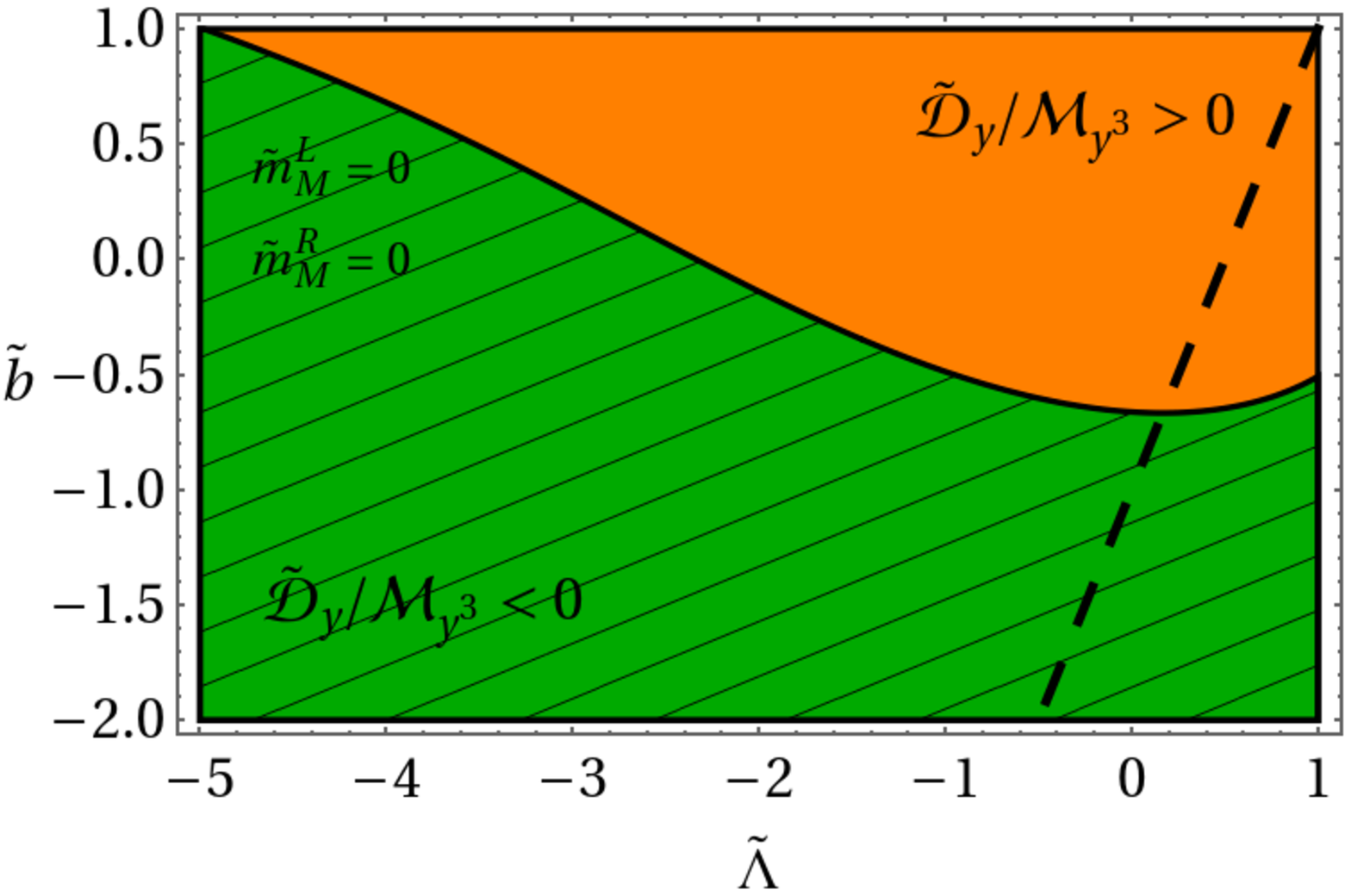} \qquad \includegraphics[width=7cm]{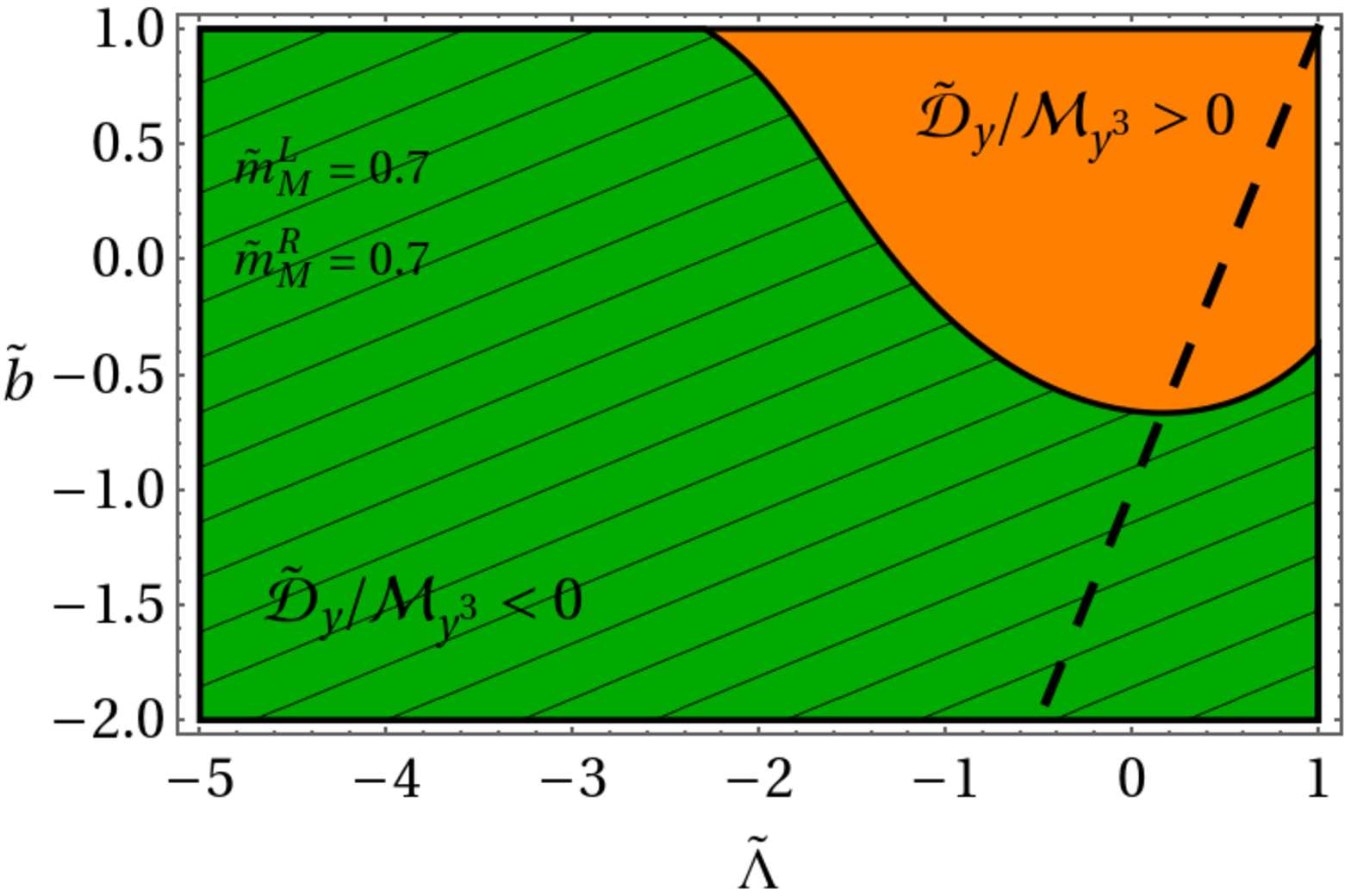} $\qquad\quad$ \\
		\includegraphics[width=7cm]{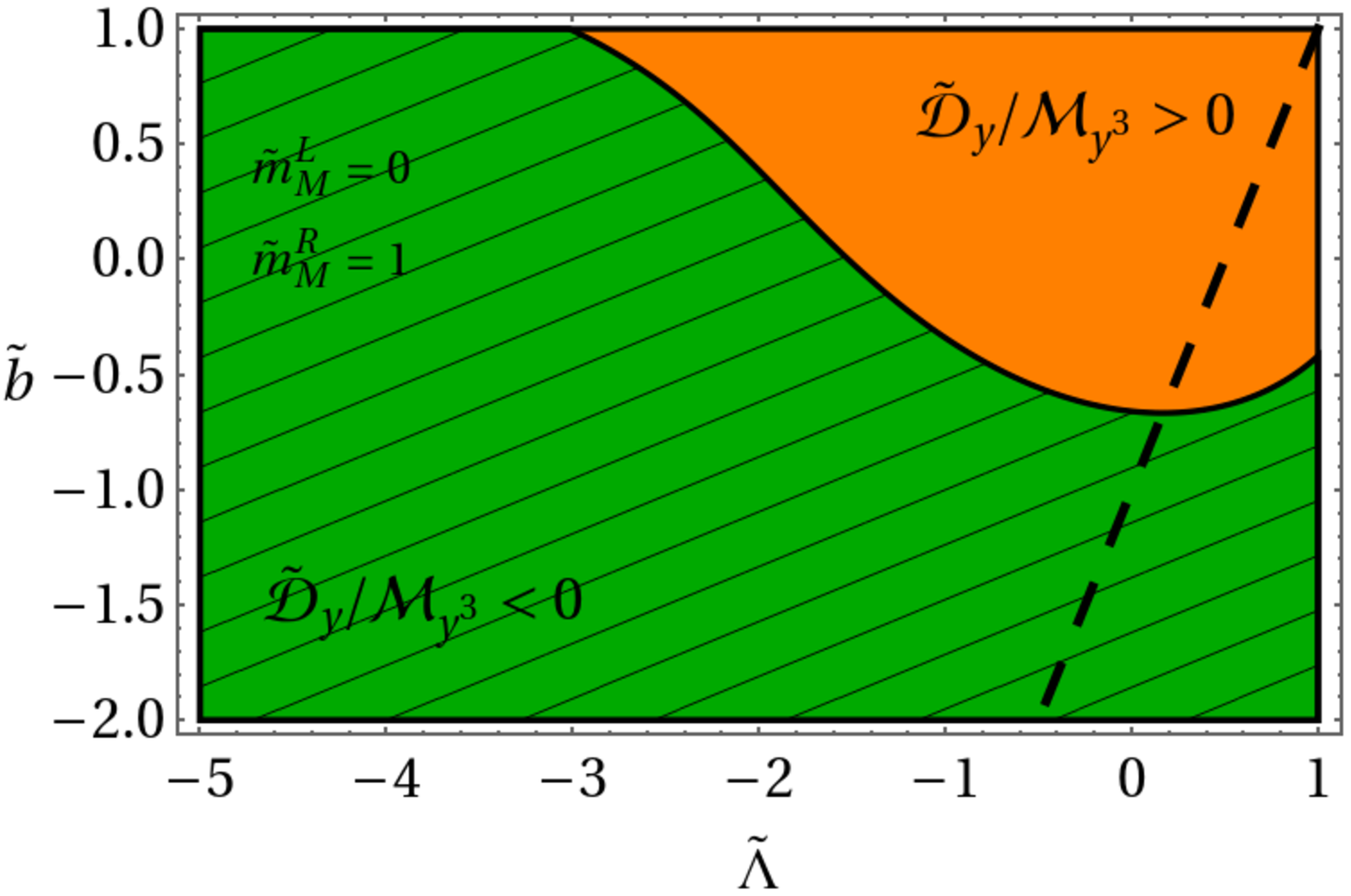} \qquad \includegraphics[width=7cm]{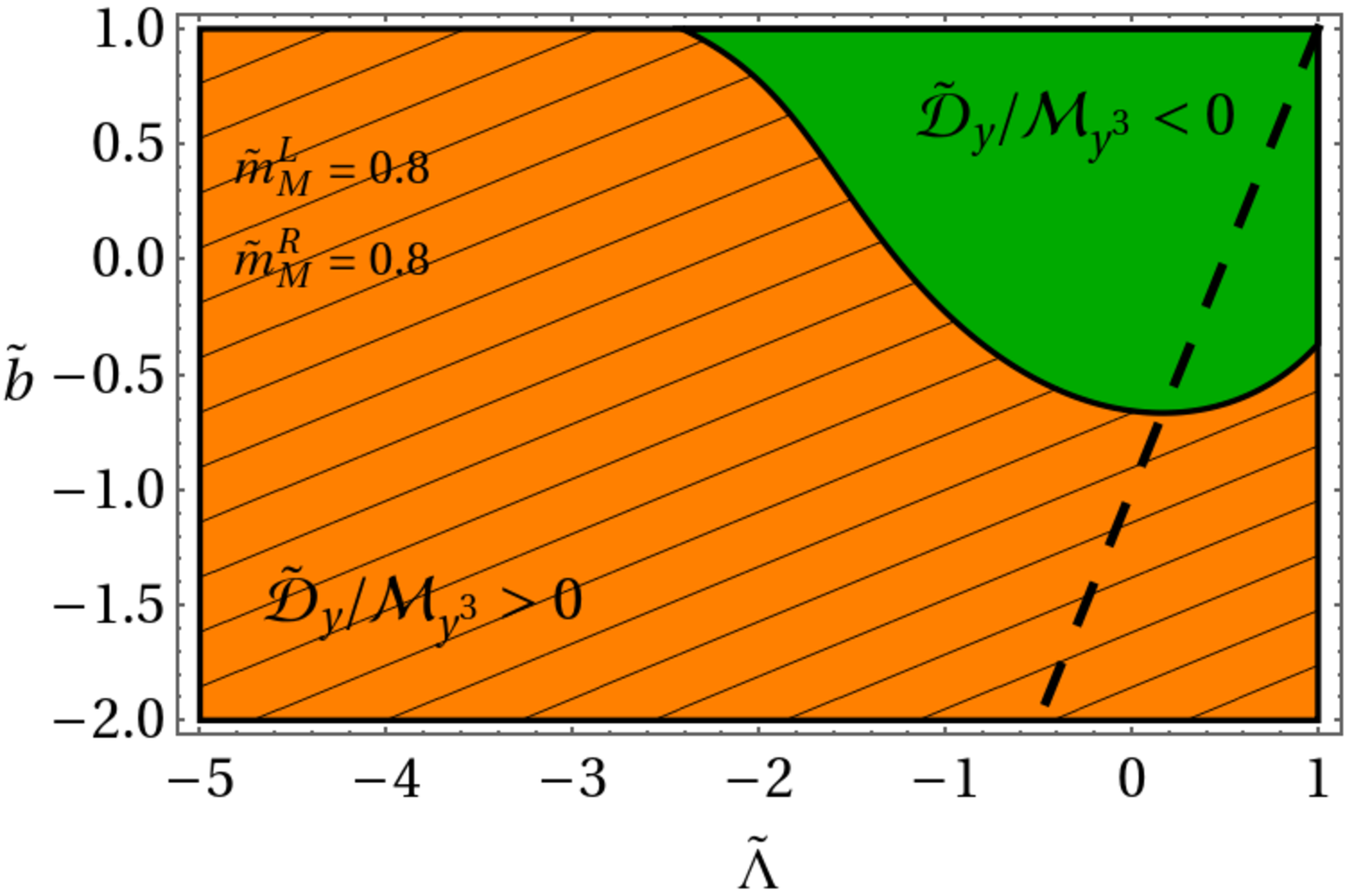} $\qquad\quad$
		\caption{Sign of $\mathcal{\tilde{D}}_y/\mathcal{M}_{y^3}$ on the ($\tilde \Lambda$, $\tilde b$) plane with $\tilde a=-0.5$. Different plots correspond to different choices of the Majorana masses. The green (darker) region correspond to $\mathcal{\tilde{D}}_y/\mathcal{M}_{y^3} < 0$, while the orange (brighter) region indicates $\mathcal{\tilde{D}}_y/\mathcal{M}_{y^3}>$. The mesh lines stand for the region constrained only by ${\tilde{\mathcal D}}_y < 0$. The dashed line indicates the pole $1 +\tilde b - 2\tilde \Lambda= 0$ coming from the $\TT$-propagator.}
		\label{Viable_Yukawa_Lambda_x_b_several_masses}
	\end{center}
\end{figure}

In order to complement the analysis for non-vanishing cosmological constant, in Fig.\,\ref{Viable_Yukawa_Lambda_x_b_several_masses}, we present the resulting sign of $\mathcal{\tilde{D}}_y/\mathcal{M}_{y^3}$ as a functions of $\tilde{\Lambda}$ and $\tilde{b}$ for some choices of the Majorana masses. These plots were generated with the choice $\tilde{a} = -0.5$ in order to avoid regions close to the pole associated with the spin-0 sector of the graviton propagator. This choice can be justified by observing that far away from the spin-0 pole the sign of $\mathcal{\tilde{D}}_y/\mathcal{M}_{y^3}$ is not affected by the variation of the parameter $\tilde{a}$. 

The collection of plots presented in this section allows us to extract important information about the viability of UV completion for the Yukawa sector. Let us start from the possibility of an asymptotically free Yukawa sector. As discussed earlier, such a scenario correspond to $\tilde{\mathcal{D}}_y < 0$ (meshed regions in Figs.\,\ref{Viable_Yukawa_a_x_b_several_masses}, \ref{Viable_Yukawa_Lambda_x_m} and \ref{Viable_Yukawa_Lambda_x_b_several_masses}) at the fixed point. The analysis on the plane $(\tilde{a} ,\tilde{b})$ (with $\tilde{\Lambda} = 0$) -- Fig.\,\ref{Viable_Yukawa_a_x_b_several_masses} -- show that variations of $\tilde{m}_M^R$ and $\tilde{m}_M^L$ have a weak effect in the sign of $\tilde{\mathcal{D}}_y$, indicating that the viability for asymptotically free UV completion as a function of higher curvature parameters is not qualitatively affected by Majorana masses.  In terms of the cosmological constant, however, the inclusion of Majorana masses plays a more interesting role. As one can see from Fig. \ref{Viable_Yukawa_Lambda_x_m}, the largest possible value of $\tilde \Lambda$ which entails an asymptotically free UV completion depends on the Majorana masses. For large values of the $\tilde{m}_M^R$ and $\tilde{m}_M^L$, the critical value for the cosmological constant becomes more negative. In particular, the case $\tilde{m}_M^R = \tilde{m}_M^L \equiv \tilde{m}_M$ (Fig.\,\ref{Viable_Yukawa_Lambda_x_m} - left) has an upper bound around $\tilde{m}_M \simeq 1.5$, in such a way that above this value we cannot have asymptotically free UV completion for the Yukawa sector. We note that the existence of such an upper bound was also verified for some -- but not all -- non-vanishing $\tilde{a}$ and $\tilde{b}$ (the value of such bound is a function of $\tilde{a}$ and $\tilde{b}$). Finally, the analysis on the plane ($\tilde{\Lambda},\tilde{b}$) -- Fig.\,\ref{Viable_Yukawa_Lambda_x_b_several_masses} -- indicates that variations of the Majorana masses have a small impact on the  qualitative behavior of the viable region for asymptotically free UV completion. 

The second scenario for UV completion of the Yukawa sector, the interacting fixed point, is characterized by $\mathcal{\tilde{D}}_y/\mathcal{M}_{y^3} < 0$. In this case, the introduction of Majorana masses play a non-trivial role. In the case of vanishing Majorana masses, the viable region for asymptotically free UV completion coincides with the regions for a viable interacting (and UV repulsive fixed point), since $\mathcal{M}_{y^3} > 0$. By turning on small values of the Majorana masses (more precisely, restricted by $\tilde{m}_M^R \tilde{m}_M^L \lesssim 1/2 $) the situation does not change, since the sign of  $\mathcal{M}_{y^3}$ remains positive. It can be seen in Figs.\,\ref{Viable_Yukawa_a_x_b_several_masses}, \ref{Viable_Yukawa_Lambda_x_m} and \ref{Viable_Yukawa_Lambda_x_b_several_masses} by noticing that the green (dark) region coincide with the meshed regions whenever $\tilde{m}_M^R \tilde{m}_M^L \lesssim 1/2$ is verified. If we increase the Majorana masses such that $\tilde{m}_M^R \tilde{m}_M^L \gtrsim 1/2$, then we find a remarkable modification. In this case the viability regions associated with each one of the scenarios for UV completion discussed here becomes complementary. In addition, the interacting fixed point turns out to be UV attractive, in contrast to what happens in the absence of Majorana masses.

An important achievement of the asymptotic safety program for quantum gravity was the prediction of the Higgs-boson mass around the measured value before the discovery of the Higgs boson~\cite{Aad:2012tfa,Chatrchyan:2012xdj}. Such a result relies on the existence of an approximated Gaussian UV repulsive fixed point for the quartic scalar coupling $\tilde{\lambda}_4$. Beyond the Planck scale quantum gravity dominates over the remaining interaction, therefore, in this regime the evolution of $\tilde{\lambda}_4$ is driven by running of the Newton coupling. The leading quantum gravity contribution to beta function associated with $\tilde{\lambda}_4$ takes the form $\beta_{\lambda_4}^{\textmd{grav.}} = \mathcal{D}_{\lambda_4} \tilde{G} \, \tilde{\lambda}_4$, where $\mathcal{D}_{\lambda_4}$ depends on the remaining gravitational parameters. For an approximated Gaussian UV repulsive fixed point the condition $\mathcal{D}_{\lambda_4} > 0$ is required at the fixed point. Several computations indicate that such a condition can be realized for simple truncations of the effective average action~\cite{Percacci:2003jz,Zanusso:2009bs,Narain:2009fy,Vacca:2010mj,Oda:2015sma,Pawlowski:2018ixd}. Below the Planck scale, on the other hand, gravity fluctuations become irrelevant and the running of $\tilde{\lambda}_4$ is driven by the Yukawa coupling. The contribution coming from the Yukawa sector to $\beta_{\lambda_4}$ is required to be negative in order to avoid instability of the quartic potential. In the framework of the SM coupled to gravity, if we set the Yukawa in accordance with the observed top-quark mass, then, the RG flow drives $\tilde{\lambda}_4$ to an IR value compatible with the observed Higgs mass~\cite{Shaposhnikov:2009pv}. 

\begin{figure}
	\begin{center}
		\includegraphics[width=7cm]{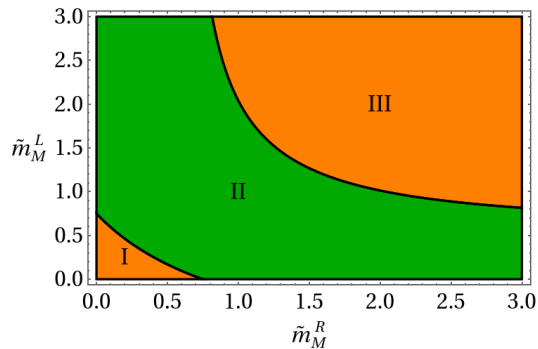}$\quad\qquad\qquad$
		\caption{Region plot showing the sign of $\mathcal{M}_{y^4}$ as a function of the Majorana masses. Regions I and III (orange/brighter) correspond to $\mathcal{M}_{y^4}<0$, while region II (green/darker) indicates $\mathcal{M}_{y^4}>0$.}
		\label{Plot-M_y^4}
	\end{center}
\end{figure}

In the framework of our toy model, we can investigate the impact of Majorana masses in a predictive scenario for the Higgs boson mass. In such a case, the beta function of the quartic coupling is given by 
\begin{align}
\beta_{\lambda_4} = ( \, 2\,\eta_\phi + \mathcal{D}_{\lambda_4} \tilde{G} \,)\, \tilde{\lambda}_4 + \mathcal{M}_{\lambda_4^2} \, \tilde{\lambda}_4^2 
+ \mathcal{M}_{y^4} \,\tilde{y}_D^4 \,.
\label{modified quartic coupling RG}
\end{align}
As we have mentioned before, a predictive scenario for the Higgs mass is subject to two conditions: i) $\mathcal{D}_{\lambda_4}$ has to be positive, which is necessary for a viable (approximated) Gaussian UV repulsive fixed point; ii) The contribution coming from the Yukawa sector has to be negative in order to avoid instabilities in the quartic potential (at least in a first approximation of the stability of the Higgs potential). The first condition is not affected by the inclusion of the Majorana masses, since $\mathcal{D}_{\lambda_4}$ depends only on the gravitational couplings. The coefficient of the $\tilde{y}_D^4$ term, on the other hand, is a function of the Majorana masses and, as a consequence, the sign of such a contribution may change by varying these parameters. Accordingly, within our truncation $\mathcal{M}_{y^4}$ is given by
\begin{align}
\mathcal{M}_{y^4} &= - \frac{1}{8\pi^2 (1 + \left(\tilde{m}_M^R)^2 \right)^3 \left(1+ (\tilde{m}_M^L)^2 \right)^3} 
\bigg[ 1 - \frac{3}{2} \,(\tilde{m}_M^R)^2 - \frac{3}{2}\, (\tilde{m}_M^L)^2 - 
\frac{1}{2}(\tilde{m}_M^R)^4 - \frac{1}{2}(\tilde{m}_M^L)^4 +
 \nonumber \\  
& \qquad\qquad -2\, \tilde{m}_M^R \tilde{m}_M^L \Big( 3 + (\tilde{m}_M^R)^2 + (\tilde{m}_M^L)^2 \Big) 
+ \frac{1}{2} (\tilde{m}_M^R \tilde{m}_M^L)^2 \Big( 4 \tilde{m}_M^R \tilde{m}_M^L + 3 (\tilde{m}_M^R)^2 + 3(\tilde{m}_M^L)^2 \Big) \bigg].
\label{My^4}
\end{align}
The sign of $\mathcal{M}_{y^4}$ is dictated by the contribution in the brackets, which is not positive definite. In Fig.\,\ref{Plot-M_y^4}, we plot the sign of $\mathcal{M}_{y^4}$ as a function of the Majorana masses. For small values of the Majorana masses, the leading contribution comes from the constant term in the brackets and, as a consequence, the sign of $\mathcal{M}_{y^4}$ remains negative (region I in Fig.\,\ref{Plot-M_y^4}). As we increase the value of the Majorana masses, those contributions coming from $\mathcal{O}(\tilde{m}_M^2)$- and $\mathcal{O}(\tilde{m}_M^4)$-terms dominates, flipping the sign of $\mathcal{M}_{y^4}$ (region II in Fig.\,\ref{Plot-M_y^4}). If we consider even larger values of the Majorana masses, the $\mathcal{O}(\tilde{m}_M^6)$ contribution becomes dominant, making $\mathcal{M}_{y^4}$ negative again (region III in Fig.\,\ref{Plot-M_y^4}). 
The $\mathcal{O}(\tilde{m}_M^6)$ contribution has the overall factor $(\tilde{m}_M^R \tilde{m}_M^L)^2$, therefore, for small values of one of the Majorana masses this contribution is suppressed, even for large values of the other mass parameter. 

Although our results are obtained from a simple toy model, we can extract some physical consequences that, in principle, can be extrapolated for more realistic scenarios. Let us start by restricting ourselves to the framework of our toy model. In this case, the viability for a predictive (toy-)Higgs boson mass can be restricted by the condition $\mathcal{M}_{y^4} < 0$. Such a condition restricts the space of allowed values for the Majorana masses which are compatible with a viable predictive scenario. 
In particular, the green region (region II) in Fig.\,\ref{Plot-M_y^4} could drive the quartic coupling to an unstable regime. 

In a more realistic scenario we should include further fermionic degrees of freedom (and also gauge fields, which are not relevant for this qualitative discussion). 
Let us assume that such additional fermions do not introduce new Majorana masses, but include extra Yukawa-Higgs interactions. 
For the sake of simplicity let us consider only one additional fermion associated with a (renormalized) Yukawa coupling denoted by $\tilde{y}^\prime$. 
In this case, the beta function for $\tilde{\lambda}_4$ receives an addition contribution like $-\frac{1}{8\pi^2}  { \tilde{y}{^\prime} }^4$.
In such a case, the viability for a predictive Higgs mass is subject to the following condition
\begin{align}
-\frac{1}{8\pi^2}  {\tilde{y}{^\prime} }^4 + \mathcal{M}_{y^4}\, \tilde{y}_D^4 < 0 \,. 
\end{align}
Numerically one can verify that $-1\leq 8\pi^2 \,\mathcal{M}_{y^4} \lesssim 0.45$ for arbitrary value of the Majorana masses. Therefore, if ${\tilde{y}{^\prime} }^4 \gtrsim 0.45 \, \tilde{y}_D^4$, the viability for a scenario with predictive Higgs mass is not affected by the introduction of a Majorana mass sector. Let $m_H^2$ be the (toy-)Higgs mass computed in the asymptotically safe quantum gravity framework neglecting the contribution of the fermionic sector associated with the Majorana masses. If we include the sector $\mathcal{M}_{y^4}\, \tilde{y}_D^4$ into the running of the quartic potential, the (toy-)Higgs mass receives an extra contribution $\delta m_H^2$. The sign of $\delta m_H^2$ is affected by the running of the Majorana masses. In particular, $\delta m_H^2 > 0$ if the flow of $\tilde{m}_M^R$ and $\tilde{m}_M^L$ lies in the regions I and III of Fig.\,\ref{Plot-M_y^4} and, $\delta m_H^2 < 0$ if the running of the Majorana masses lies in region II. The impact of the Majorana masses in the (toy-)Higgs, which can be measured $|\delta m_H^2|/(m_H^2 + \delta m_H^2 )$, is proportional to the ratio $\tilde{y}_D/\tilde{y}{^\prime}$.

\section{Summary}\label{summary section}

In this paper we have studied, using FRG techniques, a quantum-gravity matter system including a simple Higgs-Yukawa sector and Majorana masses for the fermions. We have derived the beta functions for the matter couplings and discussed their basic structures. Assuming the asymptotic safety scenario for quantum gravity, \textit{i.e.}, the running of the gravitational couplings flow towards a non-Gaussian fixed point in the UV, we investigated effects of graviton fluctuations on the running of the matter couplings. Our main findings can be summarized as follows:
\begin{itemize}
	\item Quantum gravity effects on the running of the Majorana masses:
	\begin{itemize}
		\item The fixed point characterized by $\tilde m_M^L = \tilde m_M^R = 0$ is simultaneously compatible with a non-trivial flow for the Majorana masses and UV	completion for the Yukawa coupling;
		\item The inclusion of quantum gravity effects on the running of the Majorana masses may induce an UV fixed point with $\tilde{m}_M^R \neq 0$ and/or $\tilde{m}_M^L \neq 0$.
	\end{itemize}
	\item Impact of Majorana masses on the running of Higgs-Yukawa couplings:
	\begin{itemize}
		\item Depending on the fixed point value for the Majorana masses, there are different possible scenarios for UV completion of the Yukawa coupling. The most interesting one in view point of its predictability in the IR regimes is the case where the Yukawa coupling constant has a non-trivial fixed point at which it becomes irrelevant.
		\item The inclusion of Majorana masses can affect the sign of the coefficient of the $\tilde{y}^4_D$-term and, as consequence, it may have effects on the prediction of the Higgs from asymptotically safe quantum gravity.
	\end{itemize}
\end{itemize}

We emphasize that these conclusions were obtained within a specific truncation for the effective average action and with particular choices for gauge parameters, regulators and field parametrizations (we refer to these choices as a ``scheme'' of computation). Further investigations would be necessary in order to test the stability of our results against truncation enlargement and scheme variations. Potentially relevant extensions in the pure-gravity sector involve the inclusion of terms like $a_n\,R \,(-\nabla^2)^n R$ and $b_n\,R_{\mu\nu} (-\nabla^2)^n R^{\mu\nu}$~\cite{Eichhorn:2017eht,Bosma:2019aiu,Knorr:2019atm}. In this case, an analysis analogous to the one performed in Sec.~\ref{results_2} would include additional fixed-point variables and therefore, we can expect possible deformations in our region plots. For small fixed values for these higher order couplings we expect these deformation to be small. The inclusion of additional (non-minimal) interactions in the matter sector is also necessary in order to check the stability of the fixed-point structure obtained from FRG computations for gravity-matter systems. Investigations in this direction have been performed in the last years, supporting the stability of the fixed point structure obtained within gravity-matter systems~\cite{Eichhorn:2011pc,Eichhorn:2016esv,Meibohm:2016mkp,Eichhorn:2016vvy,Eichhorn:2017eht,Eichhorn:2017sok,Eichhorn:2018nda}.

The system considered in this paper can be regarded as a toy model motivated by neutrino physics. 
An interesting feature in a system with Majorana masses is the possibility of a mass hierarchy due to the so-called seesaw mechanism.
If we assume that the neutrino is the fermion with the Majorana mass, from the fact that a neutrino mass is of order approximately $0.1$ eV, one needs a huge hierarchy between the Dirac mass and the right-handed Majorana mass, $y_D v_h\ll m_M^R $, where $v_h=246$\,GeV is the electroweak vacuum.
Note that in the framework of the Standard Model the left-handed neutrino mass is forbidden since it has a SU(2)$_L\times$U(1)$_Y$ charge.
The present analysis could be a first step to understand the compatibility of neutrinos mass mechanism in the asymptotically safe gravity scenario.
Our finding is that for a certain value of the gravitational coupling constants and the Majorana masses, there exists regions where the the Yukawa coupling constant has a non-trivial fixed point.
This indicates that within this scenario, one could identify the scale of the Majorana mass from the irrelevance of the Yukawa coupling and the observed neutrino mass.
Therefore, the possibility of future observations of the Majorana mass of neutrino could lead to consistency tests for asymptotically safe gravity.

The Yukawa coupling between a scalar boson and a fermion gives corrections to the quartic scalar coupling constant.
Through the fermion loop, the Majorana masses affect the quartic coupling constant as we have seen in \eqref{modified quartic coupling RG} and \eqref{My^4}.
By associating the scalar boson and the fermion with the Higgs boson and a neutrino, respectively, the predicted value of the Higgs boson mass from the asymptotically safe gravity scenario could be modified due to the running of the Majorana masses.
It depends on the values of the Majorana masses and whether or not the neutrino loop makes the non-trivial fixed point of the quartic coupling constant of the Higgs boson.
The current prediction from asymptotically safe quantum gravity for the Higgs boson mass is approximately $129$\,GeV, by taking up to three-loop effects on the RG equations into account within the SM~\cite{Bezrukov:2012sa}.
For a certain value of the Majorana masses, the central value of a predicted Higgs boson mass could be modified towards a smaller value.
Consequently, by taking into account effects of possible Majorana masses within the neutrino sector, the asymptotic safety scenario for quantum gravity might predict the observed Higgs boson mass $125$\,GeV.

In a future work, we will extend our analysis to a more realistic model in connection with neutrino physics. In this sense, the idea is to perform a FRG analysis within an approximation compatible with Standard Model (including Majorana masses) coupled to gravity. We will study the possibilities for UV completion on the neutrino sector and investigate the compatibility of the seesaw mechanism with the asymptotic safety scenario for quantum gravity. Depending on the structure of an UV fixed point obtained within a theory space including Majorana masses, it might be possible to predict IR values of the neutrinos Yukawa couplings and/or Majorana masses as a function of the gravitational couplings. Within this scenario, one could directly compare results from the asymptotically safe gravity scenario with the current experimental data of neutrino physics.
Together with the predictions for the Higgs and top-quark masses and so on~\cite{Shaposhnikov:2009pv,Wetterich:2016uxm,Eichhorn:2017ylw,Eichhorn:2017lry,Eichhorn:2017muy,Eichhorn:2018whv}, one can further test the consistency of the asymptotic safety scenario for quantum gravity with the observed matter degrees of freedom of our universe.

\section*{Acknowledgements}
We thank S.\,Lippoldt for valuable discussions.
The work of G.\,P.\,B. is supported by Capes under the grant no\,88881.188349/2018-01 and CNPq no.\,142049/2016-6.
The work of Y.\,H. is supported by the Advanced ERC grant SM-grav, No.\,669288. Y.\,H. thanks the hospitality of the AstroParticule et Cosmologie (APC).
The work of A.\,D.\,P was supported by the DFG through the grant Ei/1037-1.
The work of M.\,Y. is supported by the DFG Collaborative Research Centre SFB 1225 (ISOQUANT) and the Alexander von Humboldt Foundation.

\appendix

\section{Spinor on curved spacetime}
We summarize our conventions for the Dirac spinor and the Dirac gamma matrices to obtain the beta functions.
\subsection{Dirac gamma matrix and Clifford algebra}
The standard formulation of spinor in curved spacetime is given by the verbein $e_\mu^{~a}$ such that a metric is defined by $g_{\mu\nu}=e_\mu^{~a}e_\nu^{~b}\eta_{ab}$, where $\eta_{ab}=\text{diag}\,(1,-1,-1,-1)$ is the metric in flat spacetime.
In this work, we use the spin-base invariant formalism~\cite{Finster:1997gn,Weldon:2000fr} instead of the verbein formalism.
Hereafter, following the literatures~\cite{Gies:2013noa,Lippoldt:2015cea} we briefly summarize the spin-base invariant formalism.
We start with demanding that the Clifford algebra for the Dirac gamma matrices in curved spacetime is given by
\begin{align}
\{\gamma^\mu,\gamma^\nu\}=2g^{\mu\nu}{\bf 1}_{4\times 4},
\end{align}
where ${\bf 1}_{4\times 4}$ is the unity matrix in spinor space.
We require the invariance of the action for fermions under the following spin-base transformations with ${\mathcal S}\in \text{SL}\fn{4,\mathbb{C}}$: (i) $\gamma^\mu\to {\mathcal S} \gamma^\mu {\mathcal S}^{-1}$; (ii) $\psi\to {\mathcal S}\psi$; (iii) $\bar \psi \to \bar \psi{\mathcal S}$.
Here, Dirac conjugation for the Dirac spinor $\psi$ is defined by
\begin{align}
\bar\psi=\psi^\dagger {\mathfrak h},
\label{barpsi definition}
\end{align}
where ${\mathfrak h}$ is a spin metric satisfying
\begin{align}
|\text{det}\,{\mathfrak h}|=1, \qquad\qquad\qquad
{\mathfrak h}^\dagger =-{\mathfrak h}.
\label{determinant of h}
\end{align}
From the spin-base transformations (i)--(iii) and the definition \eqref{barpsi definition} we see that the spin metric is transformed under the spin-base transformations as ${\mathfrak h}\to ({\mathcal S}^{\dagger})^{-1}{\mathfrak h}{\mathcal S}^{-1}$.
In flat spacetime, one choose usually the representation ${\mathfrak h}=\gamma^0$.
In general, however, the spin metric is not equal to $\gamma^0$.
Nevertheless, it is not necessary to specify the explicit representation of ${\mathfrak h}$ for the evaluation of the Hessian derived in the next section.
The second property of the the spin metric in \eqref{determinant of h} are derived from an assumption that the mass operator of fermion is real , i.e., $(\bar\psi \psi)^*=\bar\psi \psi$.
Using this property, we see that
\begin{align}
\gamma_\mu{}^\dagger=-{\mathfrak h}\gamma_\mu {\mathfrak h}^{-1}.
\end{align}

We next define the chiral and charge conjugation operators using the Dirac gamma matrices.
The chiral operator is given by
\begin{align}
\gamma^5=-\frac{i}{4}\tilde \varepsilon_{\mu_1...\mu_4}\gamma^{\mu_1}\cdots \gamma^{\mu_4},
\end{align}
where $\tilde \varepsilon_{\mu_1...\mu_4}=\sqrt{-g}\varepsilon_{\mu_1...\mu_4}$ is the Levi-Civita tensor with $\varepsilon_{4123}=1$.
This matrix satisfies 
\begin{align}
\{ \gamma^\mu, \gamma^5\}=0, \qquad\qquad\qquad
\text{tr}\,\gamma^5=0.
\label{gamma 5 properties}
\end{align}
The charge conjugation operator is defined by
\begin{align}
C=i\gamma^2\gamma^0,
\end{align}
which satisfies
\begin{align}
C^2=-1, \qquad\qquad\qquad
C(\gamma^\mu)^TC^{-1}=-\gamma^\mu.
\label{charge conjugation property}
\end{align}

We introduce the covariant derivative
\begin{align}
D_\mu \psi =\p_\mu \psi +\Gamma_\mu \psi,\qquad\qquad\qquad
D_\mu \bar\psi =\p_\mu \bar\psi +\bar\psi \,\Gamma_\mu ,
\end{align}
where $\Gamma_\mu$ is the affine spin connection, whose spin-base transformations are give by
\begin{align}
\Gamma_\mu\to {\mathcal S}\Gamma_\mu {\mathcal S}^{-1}-(\p_\mu {\mathcal S}){\mathcal S}^{-1}.
\end{align}
Note that the covariant operator satisfies the following conditions: (i) linearity: $D_\mu (\psi_1+\psi_2)=D_\mu \psi_1+D_\mu \psi_2$: (ii) product rule: $D_\mu (\psi\bar\psi)=(D_\mu \psi)\bar\psi+\psi (D_\mu \bar\psi)$; (iii) metric compatibility: $D_\mu \bar\psi=\overline{D_\mu \psi}$; (iv) covariance: $D_\mu (\bar\psi \gamma^\nu\psi)=\nabla_\mu (\bar\psi \gamma^\nu \psi)$.
When acting the covariant derivative on the Dirac gamma matrices and the spin metric, we have
\begin{align}
D_\mu \gamma^\nu =\nabla_\mu \gamma^\nu+[\Gamma_\mu, \gamma^\nu],\qquad\qquad\qquad
D_\mu {\mathfrak h} =\p_\mu {\mathfrak h} -{\mathfrak h}\Gamma_\mu-\Gamma^\dagger {\mathfrak h}=0.
\end{align}
 
With the spin metric one can define the Dirac conjugate of a matrix $\mathcal M$ such that $\bar {\mathcal M}= {\mathfrak h}^{-1}{\mathcal M} {\mathfrak h}$, which implies $(\bar \psi {\mathcal M}\psi)^*=\bar\psi \bar{\mathcal M}\psi$.
The assumption that the kinetic term of fermion is real, namely, $\int \df^4x \sqrt{-g}\,(\bar\psi \Slash D\psi)^*=\int \df^4x \sqrt{-g}\, \bar\psi \Slash D\psi$, yields 
\begin{align}
\bar\gamma^\mu=-\gamma^\mu,
\end{align}
and 
\begin{align}
[\Delta \Gamma_\mu,\gamma^\nu] =0, \qquad\qquad\qquad
\Delta\Gamma=\Gamma_\mu-\hat\Gamma_\mu,
\label{spin torsion}
\end{align}
where $\hat \Gamma_\mu$ is an auxiliary matrix introduced in order to define the following spacetime covariant derivative 
\begin{align}
D_\text{(LC)}{}_\mu\gamma^\nu=\p_\mu \gamma^\nu+\left\{\mat{\nu\\ \mu \kappa} \right\} \gamma^\kappa=-[\hat \Gamma_\mu, \gamma^\nu], \qquad\qquad\qquad
\text{tr}\,\hat \Gamma_\mu =0,
\label{affine spin connection property}
\end{align}
with $\left\{\mat{\nu\\ \mu \kappa} \right\}$ the affine connection.
The conditions \eqref{spin torsion} are obviously satisfied by identifying $\Gamma_\mu=\hat\Gamma_\mu$.
Indeed, as discussed in \cite{Gies:2013noa}, $\Delta \Gamma_\mu$ is interpreted as the spin torsion such that
\begin{align}
D_\mu \psi=\p_\mu\psi+\hat \Gamma_\mu \psi +\Delta \Gamma_\mu \psi.
\end{align}
In this work, we do not consider the effect of the torsion, that is, we set $\Delta\Gamma_\mu=0$.
 
Finally, we mention our convention the Wick transformation from Minkowski spacetime to Euclidean spacetime.
For a vector $V_\mu\ni (x_\mu, A_\mu,\cdots)$, we define
\begin{align}
V_\text{E}{}_0=i V_0,\qquad\qquad\qquad
V_\text{E}{}_i=V_i,
\end{align}
where the subscript ``E" denotes Euclidean, and $i=1,..,4$.
The Dirac gamma matrices in Euclidean spacetime $\gamma_\text{E}{}_\mu$ are defined by
\begin{align}
\gamma_\text{E}{}_0=\gamma^0,\qquad\qquad\qquad
\gamma_\text{E}{}_i=-i\gamma^i.
\end{align}
With this definition, the Clifford algebra in flat spacetime is given as
\begin{align}
\{\gamma_\text{E}^\mu,\gamma_\text{E}^\nu\}=2\delta^{\mu\nu}{\bf 1}_{4\times 4}.
\end{align}
It should be noted that the important properties of the operators \eqref{gamma 5 properties} and \eqref{charge conjugation property} are not changed under the Wick transformation.

\subsection{Variation}
In order to calculate the Hessian for the fermion we need the formulae of the variations for the spinor field.
Let us use the local spin-based formalism with the Lorentz-symmetric gauge~\cite{Gies:2013noa,Lippoldt:2015cea}.

We consider the variation of the Dirac matrices:
\begin{align}
\gamma_\mu\fn{\bar g+h}=\bar\gamma_\mu +\frac{\p \gamma_\mu\fn{g}}{\p g_{\alpha\beta}}h_{\alpha\beta}
+\frac{1}{2}\frac{\p^2 \gamma_\mu\fn{g}}{\p g_{\alpha\beta} \p g_{\rho\sigma}}h_{\alpha\beta}h_{\rho\sigma} +\mathcal O\fn{h^3},
\end{align}
where $\bar\gamma_\mu=\gamma_\mu\fn{\bar g}$.
From the Weldon theorem~\cite{Weldon:2000fr}, the second term on the right-hand side is
\begin{align}
\frac{\p \gamma_\mu\fn{g}}{\p g_{\alpha\beta}}=\frac{1}{2}\delta^{\alpha\beta}_{\mu\nu}\gamma^\nu\fn{g}+[G^{\alpha\beta},\gamma_\mu\fn{g}],
\end{align}
with $\delta^{\alpha\beta}_{\mu\nu}=\frac{1}{2}(\delta^\alpha_\mu\delta^\beta_\nu+\delta^\alpha_\nu \delta^\beta_\mu)$.
Here, $G^{\alpha\beta}$ is a traceless tensor.
The Lorentz-symmetric gauge corresponds to a choice $G^{\alpha\beta}\fn{\bar g}=0$.
In this gauge the variations of the Dirac matrix $\gamma_\mu\fn{g}$ are evaluated as
\begin{align}
\delta \gamma_\mu= \frac{1}{2}\gamma^\nu h_{\mu\nu},\qquad\qquad\qquad
\delta^2 \gamma_{\mu} = -\frac{1}{8}h^{\lambda}_{~\mu}h_{\lambda \nu}\gamma^\nu,
\label{delgamma}
\end{align}
In the same manner, using the property \eqref{affine spin connection property} the variation of the affine spin connection is
\begin{align}
\hat \Gamma_\mu\fn{\bar g+h}
&=\hat \Gamma_\mu\fn{\bar g}+\delta \hat \Gamma_\mu+{\mathcal O}\fn{h^3}\nn
&=\hat \Gamma_\mu\fn{\bar g} + \frac{1}{8}[\gamma^\alpha, \gamma^\beta]{\bar \nabla}_\beta h_{\alpha \mu}
+\frac{1}{8}[\gamma^\alpha, \gamma^\beta]\bigg( 
-h^\rho _{~\alpha} {\bar \nabla}_\beta h_{\mu \rho}
-h^\rho_{~\beta}{\bar \nabla}_\rho h_{\mu \alpha}
-\frac{1}{2}h_{\rho \alpha}{\bar \nabla}_\mu h^{\rho}_{~\beta}
\bigg)
+{\mathcal O}\fn{h^3},
\end{align}
Then we obtain that the variation of the covariant derivative $D_\mu$ so that
\begin{align}
\delta {D}_\mu
=\frac{1}{8}[\gamma^\alpha, \gamma^\beta]{\bar \nabla}_\beta h_{\alpha \mu},\qquad\qquad
\delta^2 {D}_\mu
=\frac{1}{8}[\gamma^\alpha, \gamma^\beta]\bigg( 
-h^\rho _{~\alpha} {\bar \nabla}_\beta h_{\mu \rho}
-h^\rho_{~\beta}{\bar \nabla}_\rho h_{\mu \alpha}
-\frac{1}{2}h_{\rho \alpha}{\bar \nabla}_\mu h^{\rho}_{~\beta}
\bigg).
\label{delD2}
\end{align}

In the Lorentz-symmetric gauge, from \eqref{determinant of h}, \eqref{gamma 5 properties} and  \eqref{charge conjugation property}, one can show that ${\mathfrak h}\fn{\bar g+h}={\mathfrak h}\fn{\bar g}+\mathcal O\fn{h^3}$, $\gamma^5\fn{\bar g+h}=\gamma^5\fn{\bar g}+\mathcal O\fn{h^3}$ and $C\fn{\bar g+h}=C\fn{\bar g}+\mathcal O\fn{h^3}$, respectively.
Since the Hessians are of order of $h^2$, the variations of these operators do not contribute.

\section{Hessians}\label{list of Hessian}

In this appendix we list the Hessians computed from our truncation for the effective action given by \eqref{effective action for gravity} and \eqref{effective action for matter 1}.
Expanding $\Gamma_k$ up to second order in the fluctuation fields (both for gravity and matter sectors), namely
\begin{align}
\Gamma_k = \delta^0 \Gamma_k + \delta^1 \Gamma_k + \delta^2 \Gamma_k + \cdots \,,
\end{align}
and, keeping only those terms quadratic in the fluctuations, we can write
\begin{align}
\delta^2 \Gamma_k
=\int \df^4 x\sqrt{\bar g}\left[\,\frac{1}{2}\Phi^T \left(\Gamma_k^{(2)}\right)_{BB} \Phi
+\Psi^T\left({\Gamma_k^{(2)}}\right)_{FB}\Phi
+\Phi^T\left({\Gamma_k^{(2)}}\right)_{BF}\Psi
+\Psi^T\left({\Gamma_k^{(2)}}\right)_{FF}\Psi \,\right],
\end{align}
where we defined the fluctuation fields,
\begin{align}
\Phi := 
\begin{pmatrix}
h^\perp_{\mu\nu} & \xi_\mu & \sigma & h & \varphi
\end{pmatrix}^T
 \qquad \textmd{and} \qquad 
\Psi := \begin{pmatrix}
\delta\psi & \delta\bar\psi^T
\end{pmatrix} ^T .
\end{align}
Hereafter the background fields of the scalar boson and fermion are denoted by $\phi$ and $\psi$.
The bosonic Hessian is
\begin{align}
\left({\Gamma_k^{(2)}}\right)_{BB}=\pmat{
	\Gamma_{k,\TT}^{(2)} & 0 &  0   \\
	0 & \Gamma_{k,\xi\xi}^{(2)} & 0   \\ 
	0 & 0 &  \Gamma_k^{(2)}\big|_\text{spin\,0}
},
\end{align}
with
\begin{align}
\Gamma_k^{(2)}\big|_\text{spin\,0} =
\pmat{
	\Gamma_{k,\sigma \sigma}^{(2)} & \Gamma_{k,\sigma h}^{(2)} &  \Gamma_{k,\sigma\varphi}^{(2)} \\[5pt]
	\Gamma_{k,h\sigma}^{(2)} & \Gamma_{k,hh}^{(2)} & \Gamma_{k,h\varphi}^{(2)}\\[5pt]
	\Gamma_{k,\varphi\sigma}^{(2)} & \Gamma_{k,\varphi h}^{(2)} & \Gamma_{k,\varphi\varphi}^{(2)}
}.\label{spin 0 scalar matrix}
\end{align}

For the gravitational sector, after expansion in terms of the metric fluctuation, we perform the field redefinition $h_{\mu\nu} \mapsto Z_h^{1/2} \,\sqrt{32\pi G} \, h_{\mu\nu}$ in order to incorporate the wave function renormalization for the graviton. Such a multiplicative factor may be absorbed in the Hessians, resulting in the following explicit expressions\footnote{These expressions were computed for a maximally symmetric background metric $\bar{g}_{\mu\nu}$. In addition, the background matter field are considered to be constant (with respect to the background derivative $\bar{\nabla}_\mu$).}
\begin{subequations}
	\begin{align}\label{Hessian_TT}
	\left(\Gamma_{k,\TT}^{(2)}\right)^{\mu\nu\rho\sigma} = Z_h \,\bigg\{
	\left[\frac{}{}\! 1 + b\,\bar\Delta_{L2} - (2 a+b) \bar{R} \,\right]\left(\bar \Delta_{L2}-\frac{\bar R}{2}\right)
	+E -16\pi G\, {\mathcal I}_M 
	 \,\bigg\} \, {\bf 1}^{\mu\nu\rho\sigma} \,,
	\end{align}
	\begin{align}
	\left(\Gamma_{k,\xi\xi}^{(2)}\right)^{\mu\nu}= - 2 Z_h\,\left(\bar \Delta_{L1}-\frac{{\bar R}}{2}\right)\left[
	\, E  -\frac{1}{\alpha} \left(\bar \Delta_{L1}-\frac{{\bar R}}{2}\right) + 16 \pi G\,{\mathcal I}_M \,
	\right] {\bar g}^{\mu\nu}  \,,
	\label{T vector mode}
	\end{align}
	\begin{align}
	\Gamma_{k,\sigma \sigma}^{(2)}=
	\frac{9 \,Z_h}{8}\bigg[
	P\, {\bar \Delta}_{L0}\left({\bar \Delta}_{L0}-\frac{\bar R}{3}\right) + Q\,{\bar \Delta}_{L0}
	-\frac{2}{3}E
	-\frac{32\pi G}{3}{\mathcal I}_M
	+ \frac{1}{\alpha}\left( \bar \Delta_{L0} -\frac{\bar R}{3}\right)
	\bigg] \left( \bar \Delta_{L0} -\frac{\bar R}{3}\right) \bar \Delta_{L0}  , 
	\end{align}
	\begin{align}
	\Gamma_{k,h\sigma}^{(2)}= \Gamma_{k,\sigma h}^{(2)}
	=\frac{3\,Z_h}{8}\bigg[
	3P\left(\bar \Delta_{L0} -\frac{\bar R}{3}\right) 
	+Q
	+\frac{\beta}{\alpha}
	\bigg]\bar \Delta_{L0} \left(\bar \Delta_{L0} -\frac{\bar R}{3} \right), 
	\end{align}
	\begin{align}
	\Gamma_{k,hh}^{(2)}= \frac{9\,Z_h}{8}\bigg[
	P\left( {\bar \Delta}_{L0}-\frac{\bar R}{3}\right)^2 +Q \left( {\bar \Delta}_{L0} -\frac{\bar R}{3}\right) + \frac{2}{9}E +\frac{32\pi G}{9}{\mathcal I}_M
	+\frac{\beta^2}{9\alpha}{\bar \Delta}_{L0}
	\bigg] ,
	\end{align}
\end{subequations}
where we have defined
\begin{align}
{\mathcal I}_M=\frac{1}{2}\left(m_M^R \bar \psi^c P_R \psi +m_M^L \bar \psi^c  P_L \psi +\text{h.c.}\right) + y_D\phi \bar\psi \psi + V(\phi^2),
\end{align}
and we used the identity ${\bf 1}^{\mu\nu\rho\sigma}=({\bar g}^{\mu\rho}{\bar g}^{\nu \sigma}+{\bar g}^{\mu \sigma}{\bar g}^{\nu\rho})/2$.
Also, we used the shorthand notations,
\begin{align} \label{Defs_E_P_Q}
P=2 a +\frac{2}{3}b \,, \qquad\qquad
Q=-\frac{1}{3}+\frac{6a + 2a}{9} {\bar R}\,, \qquad\qquad
E=2\Lambda -\frac{1}{2} \bar{R} \,.
\end{align}
We also have used the following Lichnerowicz Laplacians,
\begin{subequations}
\begin{align}
{\bar \Delta}_{L0}S =-\bar \Box S \,,
\end{align}
\begin{align}
{\bar \Delta}_{L1}\xi_\mu =-\bar \Box \xi_\mu+{\bar R}_\mu^{~\alpha}\xi_\alpha \,,
\end{align}
\begin{align}
{\bar \Delta}_{L2}h_{\mu\nu} =-\bar \Box h_{\mu\nu} +{\bar R}_{\mu}^{~\alpha}h_{\alpha \nu} +{\bar R}_{\nu}^{~\alpha}h_{\mu\alpha} -{\bar R}_{\mu\alpha\nu\sigma}h^{\alpha\beta}-{\bar R}_{\mu\alpha\nu\beta}h^{\beta\alpha} \,.
\end{align}
\end{subequations}
Hereafter, the prime on $V$ denotes the derivative with respect to $\phi^2$. The Hessians for the terms with the scalar field $\phi$ are
\begin{subequations}
	\begin{align}
	\Gamma_{k,\varphi\varphi}^{(2)} = Z_\phi{\bar \Delta}_{L0} +2V'(\phi^2)+4\phi^2 V''(\phi^2) \,, 
	\end{align}
	\begin{align}
	\Gamma_{k,\sigma \varphi}^{(2)}= \Gamma_{k,\varphi\sigma}^{(2)} =0 \,,
	\end{align}
	\begin{align}
	\Gamma_{k,h \varphi}^{(2)}= \Gamma_{k,\varphi h}^{(2)} = 
	Z_h^{1/2} \,\sqrt{8\pi G}\left( y_D\bar\psi \psi + 2\phi \,V'(\phi^2) \right) \,.
	\end{align}
\end{subequations}
Note here that for $\beta=0$, the dependence of $\Gamma_k^{(h \sigma)}$ and $\Gamma_k^{(hh)}$ on $\alpha$ disappears.
Thanks to this, the mixing between $\sigma$ and $h$ does not contribute to the flow equation for $\alpha \to 0$ and then  the Hessian for the spin-0 scalar modes can be treated as a block diagonal form such that 
\begin{align}
\Gamma_k^{(2)}\big|_\text{spin\,0}=\pmat{
	\Gamma_{k,\sigma \sigma}^{(2)} & 0\\[5pt]
	0 & \Gamma_{k,\textmd{SS}}^\text{(2)} 
},
\end{align}
with the $2\times 2$ matrix for the $(h,\varphi)$ part given by
\begin{align}
\Gamma_{k,\textmd{SS}}^{(2)} =\pmat{
	\displaystyle \frac{9\,Z_h}{8}\bigg[
	P\left( {\bar \Delta}_{L0}-\frac{\bar R}{3}\right)^2 +Q \left( {\bar \Delta}_{L0} -\frac{\bar R}{3}\right) + \frac{2}{9}E +\,\frac{32\pi G}{9} {\mathcal I}_M
	\bigg] 
	& \displaystyle Z_h^{1/2} \,\sqrt{8\pi G}\left( y_D\bar\psi \psi + 2\phi \,V'(\phi^2) \right) \\[20pt]
	\displaystyle Z_h^{1/2} \,\sqrt{8\pi G}\left( y_D\bar\psi \psi + 2\phi \,V'(\phi^2) \right)
	&  Z_\phi \,{\bar \Delta}_{L0}+2V'(\phi^2)+4\phi^2 V''(\phi^2)
},
\end{align}
and the $(\sigma\sigma)$ component
\begin{align}
\Gamma_{k,\sigma\sigma}^{(2)}= \frac{9\,Z_h}{8\,\alpha}\left( \bar \Delta_{L0} -\frac{\bar R}{3}\right)^2 \bar \Delta_{L0}\,.
\end{align}
For $\alpha\to 0$ the Hessian of the transverse vector mode \eqref{T vector mode} 
\begin{align}
\left(\Gamma_{k,\xi\xi}^{(2)}\right)^{\mu\nu}
= \frac{2 \,Z_h}{\alpha} \left(\bar \Delta_{L1}-\frac{{\bar R}}{2}\right)^2 {\bar g}^{\mu\nu} 
= \frac{2 \,Z_h}{\alpha} \left(-\bar \Box-\frac{{\bar R}}{4}\right)^2 {\bar g}^{\mu\nu} .
\end{align}
Obviously, the $\xi^\mu$ and $\sigma$ modes depend on neither gravitational couplings nor the matter term $\mathcal I_M$, so that their loop effects do not contributie to the beta fucntions of the Yukawa coupling and the Majorana masses.

The Hessians for the ghost fields are also given as gravitational-coupling- and matter-independent forms:
\begin{subequations}
	\begin{align}
	\left(\Gamma_{k,\bar C^\T C^\T}^{(2)}\right)^{\mu\nu} = \left(\!-\bar \Box -\frac{\bar R}{4} \right){\bar g}^{\mu\nu},
	\end{align}
	\begin{align}
	\left(\Gamma_{k,\bar C C}^{(2)}\right) = \frac{3-\beta}{2}\left(-\bar \Box -\frac{\bar R}{3-\beta} \right)(-\bar\Box). 
	\end{align}
\end{subequations}

We show the Hessian with the fermion fields. By using \eqref{delgamma} and \eqref{delD2} we obtain the two-point function
\begin{align}
\left({\Gamma_k^{(2)}}\right)_{FF} = \begin{pmatrix}
C (m_M^R P_R + m_M^L P_L)& (Z_\psi \Slash{\bar{D}})^T - y_D \phi \\[10pt]
Z_\psi \Slash{\bar{D}} + y_D \phi & (m_M^R P_L + m_M^L P_R) C 
\end{pmatrix} .
\end{align}
The fermion propagator is, then
\begin{align}
\left({\Gamma_k^{(2)}}\right)_{FF}^{-1} = 
\begin{pmatrix}
 \Xi_2& G _1\\[5pt]
 G_2 & \Xi_1
\end{pmatrix} ,
\end{align}
where
\begin{subequations}
	\begin{align}
	G_1  = \frac{ \mathcal{A}_R \,(-Z_\psi \Slash{\bar{D}}) + y_D \phi\, \mathcal{B}}{\mathcal{A}_L \mathcal{A}_R \,(- Z_\psi^2 \Slash{\bar{D}}^2) + y_D^2 \phi^2\,\mathcal{B}^2} P_R + (R \leftrightarrow L) P_L \,, 
	\end{align}
	\begin{align}
	\Xi_1 = C \frac{-Z_\psi \Slash{\bar{D}} - y_D \phi }{- Z_\psi^2 \Slash{\bar{D}}^2 + y_D^2 \phi^2} 
	\Bigg[ \frac{m_M^L \mathcal{A}_R \,(-Z_\psi \Slash{\bar{D}}) + m_M^R  y_D \phi\, \mathcal{B} }{\mathcal{A}_L \mathcal{A}_R \,(- Z_\psi^2 \Slash{\bar{D}}^2) + y_D^2 \phi^2\,\mathcal{B}^2} P_R + (R \leftrightarrow L) P_L  \Bigg] , 
	\end{align}
	\begin{align}
	G_2 = -(\gamma^5 G_1\gamma^5)^T \qquad \textmd{and} \qquad \Xi_2 =-(\gamma^5\Xi_1\gamma^5)^T,
	\end{align}
\end{subequations}
with $\mathcal{A}_{R,L}$ and $\mathcal{B}$ being defined as
\begin{align}
\mathcal{A}_{R,L} = 1+\frac{(m_M^{R,L})^2}{-Z_\psi^2 \Slash{\bar D}^2+y_D^2\phi^2}  
\qquad \textmd{and} \qquad
\mathcal{B} = 1-\frac{m_M^Rm_M^L}{-Z_\psi^2 \Slash{\bar D}^2+y_D^2\phi^2} .
\end{align}

Using \eqref{delgamma} and \eqref{delD2}, the mixing parts are computed as
\begin{subequations}
	\begin{align}
	&\left({\Gamma_k^{(2)}}\right)_{BF} = Z_h^{1/2} \, \sqrt{32\pi G} \times
	\pmat{
		0 & 0\\
		\displaystyle -\frac{Z_\psi}{4}\left(-\bar\Box  -\frac{\bar R}{4}\right) (\bar \psi \gamma^\mu)& \displaystyle -\frac{Z_\psi}{4}\left(-\bar\Box  -\frac{\bar R}{4}\right) (\gamma^\mu \psi)^T\\[10pt]
		\displaystyle \frac{3Z_\psi}{16}\left(-\bar\Box-\frac{\bar R}{3}\right) ({\bar \nabla_\mu}\bar\psi \gamma^\mu)  & \displaystyle \frac{3Z_\psi}{16}\left(-\bar\Box-\frac{\bar R}{3}\right) (\gamma^\mu {\bar \nabla_\mu}\psi)^T \\[10pt]
		\displaystyle \frac{3Z_\psi}{16} ({\bar \nabla_\mu} \bar\psi \gamma^\mu) 
		+\frac{1}{2}\psi^T \Delta_C + \frac{y_D}{2}\phi\bar\psi
		& \displaystyle \frac{3Z_\psi}{16} ( \gamma^\mu{\bar \nabla_\mu}\psi)^T
		+\frac{1}{2}\bar\psi \tilde{\Delta}_C -\frac{y_D}{2}\phi\psi^T \\[10pt]
		y_D\bar\psi  &  -y_D\psi^T } ,
	\end{align}
	\begin{align}
	\left({\Gamma_k^{(2)}}\right)_{FB} = - \left[\left({\Gamma_k^{(2)}}\right)_{BF} \right]^T \bigg|_{\bar{\nabla}_\mu \mapsto -\bar{\nabla}_\mu} ,
	\end{align}
\end{subequations}
where $\Delta_C = C (m_M^R P_R + m_M^L P_L)$ and $\tilde{\Delta}_C = (m_M^R P_L + m_M^L P_R) C $.
Note that since the gamma matrices in the projection operators and the CP transformation are given as ones in the local Lorentz spacetime, their variations vanish.

We employ the Litim-type cutoff function \cite{Litim:2001up} by means of the following regulator
\begin{align}
\textbf{R}_k=\pmat{
	\textbf{R}_k^{BB}\fn{z} & 0 \\[5pt]
	0 & \textbf{R}_k^{FF}\fn{z} }.
\end{align}
We used the Type-I and II cutoff functions, respectively, for the bosonic and fermionic propagators. For the bosonic cutoff we can express
\begin{align}\label{Reg_BB}
\textbf{R}_k^{BB}\fn{z}= \Gamma_{k,BB}^{(2)}\fn{P_k(z)} -\Gamma_{k,BB}^{(2)}\fn{z} ,~~~\qquad\text{with $z=-\bar \Box$},
\end{align}
where $P_k(z)=z+(k^2-z)\theta(k^2-z)$,
while the fermionic one is given by
\begin{align}
\textbf{R}_k^{FF}\fn{z}= \,
\pmat{
	0 & Z_\psi{\Slash {\bar D}}^T\,r_k\fn{z/k^2} \\[5pt]
	Z_\psi{\Slash {\bar D}}\,r_k\fn{z/k^2} & 0 } ,~~~\qquad\text{with $z=- {\Slash {\bar D}}^2=-\bar \Box+{\bar R}/4$} ,
\end{align}
where the shape function is given by
\begin{align}
r_k\fn{z/k^2}=\sqrt{1+(k^2/z-1)\theta\fn{k^2/z-1}}-1.
\end{align}

\section{Structure of flow equation}\label{structure of RG flow}
The Wetterich equation for a system composed by bosons and fermions can be written in the following way
\begin{align}
\p_t \Gamma_k = \frac{1}{2} \Tr \bigg\{\left[ ( \Gamma^{(2)}_k + \textbf{R}_{k} )^{-1} \right]_{\text{BB}}\p_t \textbf{R}_{k}^{\text{BB}} \bigg\}
-  \frac{1}{2} \Tr \bigg\{\left[ (  \Gamma^{(2)}_k + \textbf{R}_{k}  )^{-1} \right]_{\Psi\Psi}\p_t \textbf{R}_{k}^{\Psi\Psi} \bigg\} .
\end{align}
Exploring the block-diagonal structure of the cut-off function (\textit{i.e.}, there is no contribution like $\textbf{R}_{k}^{\text{B} \Psi}$ or $\textbf{R}_{k}^{\text{B} \Psi}$), we can expand the traces as follows
\begin{subequations}
	\begin{align}
	\frac{1}{2}\Tr \bigg\{\left[ ( \Gamma^{(2)}_k + \textbf{R}_{k} )^{-1} \right]_{\text{BB}}\p_t \textbf{R}_{k}^{\text{BB}} \bigg\} = 
	\frac{1}{2}\Tr \left[ G_{\text{BB}}\,\p_t \textbf{R}_{k}^{\text{BB}} \right] + 
	\frac{1}{2}\Tr\left[ \Gamma^{(2)}_{\text{B} \Psi} \, G_{\Psi \Psi} \, \Gamma^{(2)}_{\Psi \text{B}} \, G_{\text{BB}} \,\p_t \textbf{R}_{k}^{\text{BB}} \,G_{\text{BB}} \right] +\cdots ,
	\end{align}
	\begin{align}
	\frac{1}{2}\Tr \bigg\{\left[ ( \Gamma^{(2)}_k + \textbf{R}_{k} )^{-1} \right]_{\Psi\Psi}\p_t \textbf{R}_{k}^{\Psi\Psi} \bigg\} = 
	\frac{1}{2}\Tr \left[ G_{\Psi\Psi}\,\p_t \textbf{R}_{k}^{\Psi\Psi} \right] + 
	\frac{1}{2}\Tr\left[ \Gamma^{(2)}_{\Psi \text{B}} \, G_{\text{BB}} \, \Gamma^{(2)}_{\text{B} \Psi} \, G_{\Psi\Psi} \,\p_t \textbf{R}_{k}^{\Psi\Psi} \,G_{\Psi\Psi} \right] +\cdots ,
	\end{align}
\end{subequations}
where we have defined $G_{\text{BB}}  = (\, \Gamma^{(2)}_{k,\text{BB}} + \textbf{R}_k^{\text{BB}} \,)^{-1}$ and $G_{\Psi\Psi}  = (\Gamma^{(2)}_{\Psi\Psi} + \textbf{R}_k^{\Psi\Psi} )^{-1}$. It is important to emphasize that the first two terms of these expansion are sufficient for the purposes of this paper, therefore, we can truncate expansion without any loss. After York decomposition, the first trace in the bosonic sector can be expressed as\footnote{Note that we are including the Faddeev-Popov ghosts field as part of the bosonic multiplet, despite of its anti-commuting nature. Moreover, the Jacobian contribution was also include in the bosonic trace.}
\begin{align}
\frac{1}{2}\Tr \left[ G_{\text{BB}}\,\p_t \textbf{R}_{k}^{\text{BB}} \right] &=  \frac{1}{2}\Tr_{(\text{2TT})} \left[ G_{\TT} \, \p_t \textbf{R}_k^{\TT}  \right]
+  \frac{1}{2} \Tr'_{(\text{1\T})} \left[ G_{\xi\xi} \, \p_t \textbf{R}_k^{\xi\xi}  \right] 
+ \frac{1}{2} \Tr''_{(0)} \left[ G_{\text {spin-0}} \, \p_t \textbf{R}_k^{\text {spin-0}}  \right] +  \nonumber \\
&+ \frac{1}{2}\sum_{j=0}^1 \left[ G_{\text{SS}}(\lambda_j)\, \p_t \textbf{R}_{k}^{\text{SS}}(\lambda_j)  \right] 
-\Tr_{(\text{1T})} \left[ G_{\bar C^T \!C^T} \,\p_t \textbf{R}_k^{\bar C^T \!C^T} \right]   
- \Tr_{(0)}' \left[ G_{\bar C^L \!C^L} \,\p_t \textbf{R}_k^{\bar C^L \!C^L} \right] +  \nonumber \\
&-\frac{1}{2} \Tr^\prime_{(1\text{T})} \Bigg[ \frac{\p_t P_k}{P_k -\bar R/4} \Bigg]
-\frac{1}{2} \Tr_{(0)}'' \Bigg[ \frac{\p_t P_k \left( 2\,P_k - \bar R /3 \right)}{ P_k\, \left(P_k- \bar R/3\right) } \Bigg]
+   \Tr_{(0)}' \bigg[ \frac{\p_t P_k}{P_k} \bigg] ,
\end{align}
where $G_{ij}  = \left[(\, \Gamma^{(2)}_{k} + \textbf{R}_k \, )^{-1} \right]_{ij}$ for every pair $(i,j)$. It is convenient to define a short notation
\begin{align}
\frac{1}{2}\Tr \left[ G_{\text{BB}}\,\p_t \textbf{R}_{k}^{\text{BB}} \right] = \mathcal{T}_{(2\TT)} + \mathcal{T}_{(1\T)} + \mathcal{T}_{(0)}\,, 
\end{align}
where $\mathcal{T}_{(s)}$ is defined by the sum of all contributions with spin-$s$. As discussed in the previous section, a remarkable simplification is achieved by the gauge choice $\alpha \to 0$ and $\beta = 0$. In this case, the contributions coming from the spin-1 and spin-0 can be summarized in the following expressions
\begin{align}
\mathcal{T}_{(1T)} = -\frac{1}{2}  \Tr_{(\text{1T})}' \Bigg[ \frac{\p_t {P}_{k} }{ P_k- \bar R/4 } \Bigg] 
+ \frac{\eta_h}{2}  \, \Tr_{(\text{1T})}' \Bigg[ \frac{ \left(  P_k- \bar R/4 \right)^2  - \left(  -\bar \Box - \bar R/4 \right)^2}{ \left(  P_k- \bar R/4 \right)^2 } \Bigg]
- \frac{\p_t {P}_{k}(\lambda_1) }{ P_k(\lambda_1) - \bar R/4 } 
:= \Delta \mathcal{T}_{(1)},
\end{align}
and
\begin{align}
\mathcal{T}_{(0)} = \frac{1}{2} \Tr_{(0)} \left[ G_{\text{SS}} \, \p_t \textbf{R}_k^{\text{SS}}  \right] + \Delta \mathcal{T}_{(0)} ,
\end{align}
where
\begin{align}
\Delta\mathcal{T}_{(0)} := -\frac{1}{2}  \Tr_{(0)}'' \Bigg[ \frac{\p_t {P}_{k} }{ P_k- \bar R/3 }\Bigg]+ \frac{\eta_h}{2}   \, \Tr_{(0)}'' \Bigg[ \frac{ \left(  P_k- \bar R/3 \right)^2 P_k  - \left(  -\bar \Box - \bar R/3 \right)^2(-\bar \Box)}{ \left( P_k- \bar R/3 \right)^2 P_k } \Bigg]  - \frac{\p_t {P}_{k}(\lambda_1) }{ P_k(\lambda_1)- \bar R/3 }  .
\end{align}
Note that SS denotes $\varphi$ and $h$.

The aforementioned choice for the gauge parameter also simplifies those contributions to the trace involving fermions-boson  mixing. In fact, it is not difficult to conclude that, in this case, the contribution coming from the bosonic sector are those associated with the spin-0 fields $h$ and $\varphi$ (i.e., from the sub-sector $\Gamma_{\text{SS}}^{(2)}$). Therefore, the mixed contributions to the flow equation can be cast in the following way
\begin{align}
\mathcal{T}_{\text{mixed}}  :=&\,\,
\frac{1}{2}\Tr\left[ \Gamma^{(2)}_{\text{B} \Psi} \, G_{\Psi \Psi} \, \Gamma^{(2)}_{\Psi \text{B}} \, G_{\text{BB}} \,\p_t \textbf{R}_{k}^{\text{BB}} \,G_{\text{BB}}\right]
- \frac{1}{2}\Tr\left[ \Gamma^{(2)}_{\Psi \text{B}} \, G_{\text{BB}} \, \Gamma^{(2)}_{\text{B} \Psi} \, G_{\Psi\Psi} \,\p_t \textbf{R}_{k}^{\Psi\Psi} \,G_{\Psi\Psi} \right] \nonumber \\[2ex]
=&\,\, \frac{1}{2}\Tr\left[ \Gamma^{(2)}_{\text{S} \Psi} \, G_{\Psi \Psi} \, \Gamma^{(2)}_{\Psi \text{S}} \, G_{\text{SS}} \,\p_t \textbf{R}_{k}^{\text{SS}} \,G_{\text{SS}}\right] 
- \frac{1}{2}\Tr\left[ \Gamma^{(2)}_{\Psi \text{S}} \, G_{\text{SS}} \, \Gamma^{(2)}_{\text{S} \Psi} \, G_{\Psi\Psi} \,\p_t \textbf{R}_{k}^{\Psi\Psi} \,G_{\Psi\Psi} \right] .
\end{align} 
For the pure fermionic trace we denote
\begin{align}
\mathcal{T}_{(1/2)} = - \frac{1}{2}\Tr \left[ G_{\Psi\Psi}\,\p_t \textbf{R}_{k}^{\Psi\Psi} \right].
\end{align}
Putting all pieces together, we can rewrite the flow equation in the compact form
\begin{align}
\p_t \Gamma_k =  \mathcal{T}_{(\text{2TT})} + \mathcal{T}_{(\text{1T})} + \mathcal{T}_{(0)}  + \mathcal{T}_{(1/2)} + \mathcal{T}_{\text{mixed}}.
\end{align}

To illustrate the evaluation of these kernels ${\mathcal T}_i$, let us demonstrate a detailed part of the computation of the gravity contribution to the running of matter couplings.\footnote{Detailed computations for the running of the gravitational couplings have been previously discussed in the literature, e.g., \cite{Percacci:2017fkn}.} For the sake of simplicity we focus on the contribution coming from the TT sector. Computations for different sectors can be performed analogously. Since we are dealing with contributions to the running of matter couplings, it is convenient to set the background metric to be flat, i.e., $\bar g_{\mu\nu} = \delta_{\mu\nu}$. Using \eqref{Hessian_TT}, \eqref{Defs_E_P_Q} and \eqref{Reg_BB}, we can write
\begin{align}
[\mathcal{T}_{(\text{2TT})}]_{\text{Proj. Flat}} &= \frac{1}{2} \Tr_{(2\TT)}\bigg[
\frac{\p_t [ Z_h( 1 + b\,P_k ) \,P_k - Z_h( 1 - b\,\bar \Box ) \,(-\bar \Box)]}{
	Z_h[( 1 + b\,P_k ) \,P_k +2\Lambda -16\pi G\, {\mathcal I}_M ] } \bigg] \nonumber \\
&= \frac{1}{2} \Tr_{(2\TT)}\bigg[
\frac{\p_t [ Z_h( 1 + b\,P_k ) \,P_k - Z_h( 1 - b\,\bar \Box ) \,(-\bar \Box)]}{Z_h [( 1 + b\,P_k ) \,P_k
	+2\Lambda ]} \bigg( 1 + \frac{16\pi G\, {\mathcal I}_M}{( 1 + b\,P_k ) \,P_k
	+2\Lambda}\bigg) + \cdots \bigg] .
\end{align}
In order to project on the contributions to the running of matter couplings we select only those terms linear in ${\mathcal I}_M$, namely 
\begin{align}
[\mathcal{T}_{(\text{2TT})}]_{\text{Proj. Matter}} = 
\frac{1}{2}\Tr_{(2\TT)}\bigg[16\pi G \, {\mathcal I}_M \,\times \,
\frac{\p_t [ Z_h( 1 + b\,P_k ) \,P_k - Z_h( 1 - b\,\bar \Box ) \,(-\bar \Box)]}{ 
Z_h\,[( 1 + b\,P_k ) \,P_k + 2\Lambda]^2 } \bigg] .
\end{align}
Setting the matter fields (contained in ${\mathcal I}_M$) to constant configurations the trace can be expressed in terms of a momentum integral. Performing the momentum integral and expressing the result in terms of dimensionless coupling and fields yield
\begin{align}
[\mathcal{T}_{(\text{2TT})}]_{\text{Proj. Matter}} &= \frac{5\,\tilde{G}}{4\pi}
\bigg(1-\frac{\eta_h}{6}\bigg) \frac{(2+3\tilde b)}{(1+\tilde b-2\tilde \Lambda)^2} \,\nn 
&\,\,\times \int d^4x\, \sqrt{\bar{g}} \,
\bigg[\frac{k}{2}\left(\tilde m_M^R  {\tilde{\bar \psi}}^{\,c} P_{R} \tilde{\psi} +
\tilde m_M^L  {\tilde{\bar \psi}}^{\,c} P_{L} \tilde{\psi}  +\text{h.c.}\right) + y_D\phi \bar\psi \psi + V(\phi^2) \bigg].
\end{align}
Projecting on the Majorana-right (left) mass term, we find
\begin{align}
[\mathcal{T}_{(\text{2TT})}]_{\text{Proj. }\tilde{m}_M^{R(L)}} =
\frac{5\,k \,\tilde{G} \,\tilde{m}_M^{R(L)}}{8\pi}
\bigg(1-\frac{\eta_h}{6}\bigg) \frac{(2+3\tilde b)}{(1+\tilde b-2\tilde \Lambda)^2}  
\int d^4x\, \sqrt{\bar{g}} \,\left( {\tilde{\bar \psi}}^{\,c} P_{R(L)} \tilde{\psi}  + \textmd{h.c.} \right).
\end{align}
Using \eqref{Flow_Majorana} it follows that the TT-contribution to the Majorana-right (left) beta function is given by 
\begin{align}
\beta_{m_M^{R(L)}}\Big|_{(2\TT)} = 
\frac{5 \,\tilde{G} \,\tilde{m}_M^{R(L)}}{4\pi}
\bigg(1-\frac{\eta_h}{6}\bigg) \frac{(2+3\tilde b)}{(1+\tilde b-2\tilde \Lambda)^2} .
\end{align}
Similar computations can be performed in order to extract the running of the remaining sectors.

\section{Beta functions}\label{Explicit form of beta functions}
We list the explicit forms of the beta functions in four dimensional spacetime.
To this end, we define a threshold functions:
\begin{align}
\hspace{-.11cm}{\mathcal I}_{n_T,n_g;n_L,n_R;n_s} \!\!=\!\!\!\!\!\!\quad\left( 1-2\tilde \Lambda+{\tilde b} \right)^{-n_T}\!\left( 3-4\tilde \Lambda -18 \tilde a -6\tilde b  \right)^{-n_g}\!\Big(1\!+\!({\tilde m}_M^L)^2 \Big)^{-n_L}\!\Big(1\!+\! ({\tilde m}_M^R)^2 \Big)^{-n_R}\!\left( 1\!+\!2\tilde \lambda_2\right)^{-n_s}.
\label{threshold function}
\end{align}
We define the graviton anomalous dimensions as $\eta_h = - Z_h^{-1} \p_t Z_h$. It is also useful to define
\begin{align}
\eta_f= \eta_h-\frac{\p_t(1 -9{\tilde a} - 3\tilde b)}{1 - 9{\tilde a} -3\tilde b} 
\qquad\quad \textmd{and} \quad\qquad\eta_g=\eta_h- \frac{ \p_t( 5  -48{\tilde a} - 16\tilde b)}{5  -48{\tilde a} - 16\tilde b } \,.
\end{align}

Before performing  the explicit evaluation of the beta functions, we analytically investigate their structure 
in order to gain some insights at particular limits.
We start by looking at the structure of  the spin 2 and 0 graviton propagators\footnote{These expressions are obtained by expanding the full two-point function $\left(\Gamma^{(2)}_{k}\right)^{-1}$ in powers of $\bar R$ and retaining the $\bar R$-independent part.}, namely
\begin{align}
\frac{1}{1-2\tilde \Lambda+\tilde b}\, , \qquad \qquad\qquad
\frac{1}{3 -4\tilde \Lambda -18\tilde a -6\tilde b} \,.
\label{graviton propagators}
\end{align}
Around the poles of these propagators the graviton fluctuations become strong.
When the fixed point value lies near the poles, the critical exponent tends to be large.
However, such a case is not acceptable as a reliable result since before one is dealing with very stable and convergent truncations, the fixed point values and thus the proximity to the propagator poles is truncation-dependent. This prevents robust conclusions due to the artificial enhancement of gravitational fluctuations.

\begin{figure}[htb!]
	\begin{center}
		\includegraphics[width=11cm]{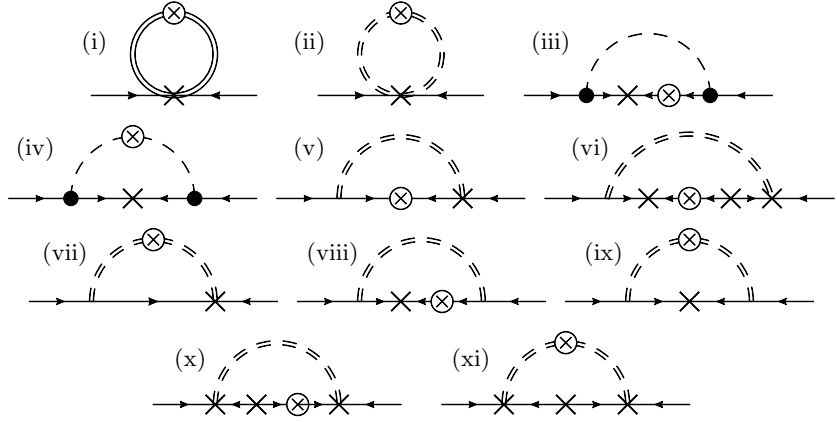}
		\put(-285,140){(i)}
		\put(-205,140){(ii)}
		\put(-115,140){(iii)}
		\put(-310,100){(iv)}
		\put(-205,100){(v)}
		\put(-100,100){(vi)}
		\put(-300,60){(vii)}
		\put(-200,60){(viii)}
		\put(-95,60){(ix)}
		\put(-250,20){(x)}
		\put(-145,20){(xi)}
		\caption{Diagrams contributing to the left-handed Majorana mass in the Landau gauge ($\alpha\to 0$) with $\beta=0$.
			The single-solid and broken lines exhibit fermion and scalar-boson ($\phi$), respectively, while the double-solid and broken lines are spin 2 ($h_{\mu\nu}^\mathrm{TT}$) and spin 0 ($h={\bar g}_{\mu\nu}h^{\mu\nu}$) gravitons, respectively.
			The cross in circle denotes the cutoff insertion into the propagators, i.e. $\otimes=\p_t R_k$. The The cross and black-point stand for the Majorana masses and the Yukawa coupling, respectively.}
		\label{Feynman diagrams of Majorana beta functions}
	\end{center}
\end{figure}

The dimensionless beta function of the left-handed Majorana mass is
\begin{align}
\beta_{m_M^L}=& -(1-\eta_\psi){\tilde m}_M^L + \bigg\{ ({\rm i}) +({\rm ii})+\big[({\rm iii})+({\rm iv})\big]+\big[({\rm v})+({\rm vi})+({\rm vii})\big]+\big[({\rm viii})+({\rm ix})\big] +\big[ ({\rm x})+({\rm xi})\big]\bigg\},
\label{Left-handed Majorana mass beta function}
\end{align}
where
\begin{subequations}
\begin{align}
({\rm i})&=\frac{5\tilde m_M^L \tilde G}{4\pi}\left(1 - \frac{\eta_h}{6}\right)\left(2 + 3\tilde b \right){\mathcal I}_{2,0;0,0;0}\,,
\label{TT contribution in majorana}
 \\
({\rm ii})&=\frac{3\tilde m_M^L \tilde G}{\pi}\left(1 -\frac{\eta_f}{6} \right)\left(  1  -9{\tilde a} - 3\tilde b \right){\mathcal I}_{0,2;0,0;0}\,,\\
({\rm iii})+({\rm iv})&=\frac{\tilde y_D^2{\tilde m}_M^R }{16\pi^2}\Bigg[ \left(1-\frac{\eta_\psi}{5} \right) 
{\mathcal I}_{0,0;2,0;1}
+\left(1-\frac{\eta_\phi}{6} \right) \Big({\mathcal I}_{0,0;1,0;2} \Big)\Bigg],\\
({\rm v})+({\rm vi})+({\rm vii})&=-\frac{6\tilde m_M^L \tilde G}{5\pi}\Bigg[
\left(1-(m_{M}^L)^2\right)\left(1-\frac{\eta_\psi}{6}\right) {\mathcal I}_{0,1;2,0;0}\nn
&\qquad
+\frac{6}{9}\left\{ 4\left( 1-\frac{\eta_f}{7}\right) \left( 1  -9{\tilde a} - 3\tilde b \right)+\left( 1-\frac{\eta_g}{7}\right) \left( 5  -48 {\tilde a}-16\tilde b \right)  \right\} {\mathcal I}_{0,2;1,0;0} 
\Bigg],\\
({\rm viii})+({\rm ix})&=\frac{3\tilde m_M^L \tilde G}{8\pi}\Bigg[
\left(1-\frac{\eta_\psi}{7}\right) {\mathcal I}_{0,1;2,0;0}
	+ \frac{6}{10}\left( 1-\frac{\eta_g}{8}\right) \left( 5  -48{\tilde a}-16\tilde b \right) {\mathcal I}_{0,2;1,0;0}
\Bigg],\\
({\rm x})+({\rm xi})&=-\frac{4({\tilde m}_M^L)^3 \tilde G}{\pi}\Bigg[ \left(1-\frac{\eta_\psi}{5}\right){\mathcal I}_{0,1;2,0;0}
+3\left(1 -\frac{\eta_f}{6}\right) \left(1  -9{\tilde a} -3\tilde b \right) {\mathcal I}_{0,2;1,0;0} \Bigg].
\end{align}
\end{subequations}
Each contribution to the left-handed Majorana mass corresponds to the diagrams shown in Fig.~\ref{Feynman diagrams of Majorana beta functions}.
The right-handed one is obtained by the replacing ${\tilde m}_M^L \leftrightarrow {\tilde m}_M^R$. 

\begin{figure}
\begin{center}
\includegraphics[width=16cm]{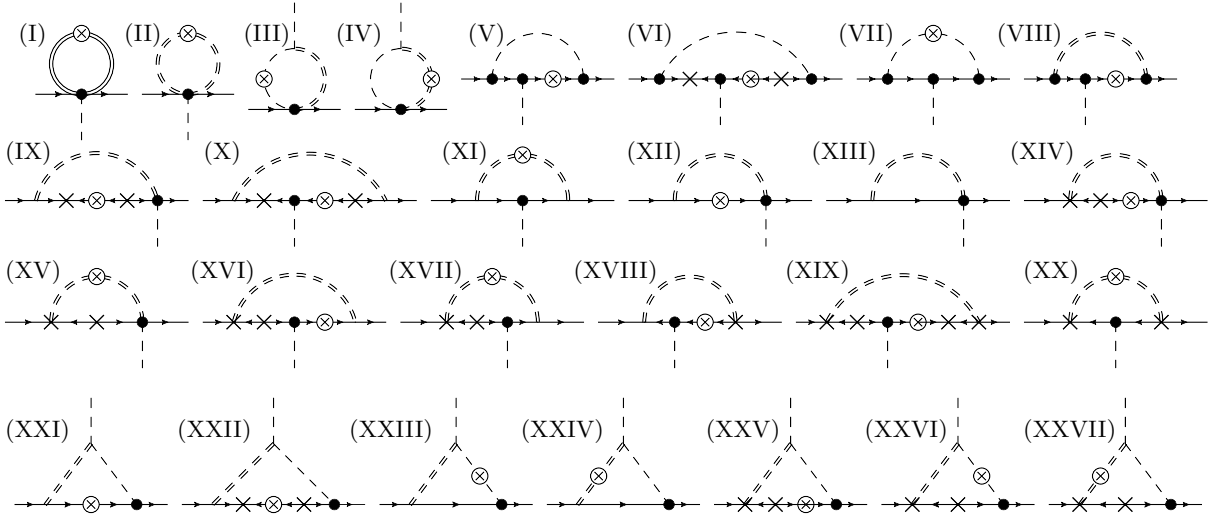}
\put(-450,180){(I)}
\put(-410,180){(II)}
\put(-365,180){(III)}
\put(-330,180){(IV)}
\put(-280,180){(V)}
\put(-220,180){(VI)}
\put(-140,180){(VII)}
\put(-80,180){(VIII)}
\put(-455,135){(IX)}
\put(-380,135){(X)}
\put(-290,135){(XI)}
\put(-220,135){(XII)}
\put(-150,135){(XIII)}
\put(-75,135){(XIV)}
\put(-455,90){(XV)}
\put(-385,90){(XVI)}
\put(-310,90){(XVII)}
\put(-240,90){(XVIII)}
\put(-160,90){(XIX)}
\put(-70,90){(XX)}
\put(-455,30){(XXI)}
\put(-390,30){(XXII)}
\put(-325,30){(XXIII)}
\put(-260,30){(XXIV)}
\put(-190,30){(XXV)}
\put(-130,30){(XXVI)}
\put(-70,30){(XXVII)}
\caption{Diagrams contributing to the Yukawa coupling constant in the Landau gauge ($\alpha\to 0$) with $\beta=0$.
The single-solid and broken lines exhibit fermion and scalar-boson ($\phi$), respectively, while the double-solid and broken lines are spin 2 ($h_{\mu\nu}^\mathrm{TT}$) and spin 0 ($h={\bar g}_{\mu\nu}h^{\mu\nu}$) gravitons, respectively.
The cross in circle denotes the cutoff insertion into the propagators, i.e. $\otimes=\p_t R_k$. The cross and black-point stand for the Majorana mass and the Yukawa coupling, respectively.}
\label{Feynman diagrams of Yukawa beta functions}
\end{center}
\end{figure}

The dimensionless beta function of the Yukawa coupling constant is 
\begin{align}
\label{eq:beta_yukawa}
\beta_y=&
\left(\frac{\eta_\phi}{2}+ \eta_\psi \right)\tilde y_D +\bigg\{({\rm I}) + \cdots + \big[  ({\rm XXV})+({\rm XXVI})+({\rm XXVII})\big]\bigg\},
\end{align}
where
\begin{subequations}
\begin{align}
({\rm I})&=\frac{5\tilde y_D \tilde G}{4\pi}\left( 1-\frac{\eta_h}{6} \right)\left(2+3\tilde b\right){\mathcal I}_{2,0;0,0;0}\, ,\\
({\rm II})&=\frac{3\tilde y_D \tilde G}{\pi} \left( 1 - \frac{\eta_f}{6} \right) \left( 1 -9{\tilde a}-3\tilde b \right){\mathcal I}_{0,2;0,0;0}\, ,\\
({\rm III})+({\rm IV})&=-\frac{8\tilde y_D \tilde \lambda_2 \tilde G}{\pi}\Bigg[
\left(1-\frac{\eta_\phi}{6} \right){\mathcal I}_{0,1;0,0;2} + 3 \left(1-\frac{\eta_f}{6}\right)\Big( 1  -9{\tilde a} -3\tilde b \Big) {\mathcal I}_{0,2;0,0;1}
\Bigg],\\
({\rm V})+({\rm VI})+({\rm VII})&=\frac{\tilde y_D^3}{16\pi^2}\Bigg[ \left(1-\frac{\eta_\psi}{5}\right) \Big(1-{\tilde m}_M^R{\tilde m}_M^L \Big\{ 2+({\tilde m}_M^R)^2+({\tilde m}_M^L)^2 +{\tilde m}_M^R{\tilde m}_M^L\Big\}\Big){\mathcal I}_{0,0;2,2;1} +\nn
&\quad
+\left( 1-\frac{\eta_\phi}{6}\right) 
(1-{\tilde m}_M^L{\tilde m}_M^R){\mathcal I}_{0,0;1,1;2} \Bigg],\\
({\rm VIII})+({\rm IX})+({\rm X})&=\frac{3\tilde y_D \tilde G}{8\pi}\Bigg[ \left(1-\frac{\eta_\psi}{7}\right) 
\Big(1-{\tilde m}_M^R{\tilde m}_M^L \Big\{ 2+({\tilde m}_M^R)^2+({\tilde m}_M^L)^2 +{\tilde m}_M^R{\tilde m}_M^L\Big\}\Big){\mathcal I}_{0,1;2,2;0} + \nn
&\quad
+\frac{3}{5}\left( 1 -\frac{\eta_g}{8} \right) \left(5  -48{\tilde a } -16\tilde b \right)  (1-{\tilde m}_M^L{\tilde m}_M^R){\mathcal I}_{0,2;1,1;0} \Bigg],\\
({\rm XI})+({\rm XII})+({\rm XIII})&=- \frac{3\tilde y_D \tilde G}{5\pi}\Bigg[ \left(1-\frac{\eta_\psi}{6}\right)\Big(\left(1-(\tilde m_M^L)^2 \right){\mathcal I}_{0,1;2,0;0} + \left(1-(\tilde m_M^R)^2 \right){\mathcal I}_{0,1;0,2;0} \Big) + \nn
&\quad
+\frac{2}{3} \left\{ 4\left( 1-\frac{\eta_f}{7}\right) \left( 1 -9{\tilde a} -3\tilde b \right)+\left( 1-\frac{\eta_g}{7}\right) \left( 5  -48{\tilde a} -16\tilde b\right)  \right\} 
{\mathcal I}_{0,2;1,0;0} + \nn
&\quad
+\frac{2}{3} \left\{ 4\left( 1-\frac{\eta_f}{7}\right) \left( 1 -9{\tilde a} -3\tilde b \right)+\left( 1-\frac{\eta_g}{7}\right) \left( 5  -48{\tilde a} -16\tilde b\right)  \right\} {\mathcal I}_{0,2;0,1;0} 
\Bigg] ,\\
({\rm XIV})+({\rm XV})&=-\frac{4\tilde y_D \tilde G}{\pi}\Bigg[\left(1-\frac{\eta_\psi}{5}\right) \Big(({\tilde m}_M^L)^2 {\mathcal I}_{0,1;2,0;0} +({\tilde m}_M^R)^2 {\mathcal I}_{0,1;0,2;0}\Big) + \nn
&\quad
+3\left(1 -\frac{\eta_f}{6}\right)\left(1 -9{\tilde a} - 3\tilde b \right)  \Big( ({\tilde m}_M^L)^2{\mathcal I}_{0,2;1,0;0}+({\tilde m}_M^R)^2{\mathcal I}_{0,2;0,1;0}  \Big)\Bigg],\\
({\rm XVI})+({\rm XVII})&=\frac{3\tilde y_D\Big( {\tilde m}_M^L+{\tilde m}_M^R\Big)^2 \tilde G}{5\pi}\Bigg[ \left(1-\frac{\eta_\psi}{6}\right)\Big(3+({\tilde m}_M^L)^2+({\tilde m}_M^R)^2 - ({\tilde m}_M^L)^2({\tilde m}_M^R)^2\Big) {\mathcal I}_{0,1;2,2;0} +\nn
&\quad
+\frac{2}{3}\left\{ 4\left( 1-\frac{\eta_f}{7}\right) \left( 1 -9 {\tilde a} - 3\tilde b \right)+\left( 1-\frac{\eta_g}{7}\right) \left( 5 -48 {\tilde a} -16\tilde b \right)  \right\}{\mathcal I}_{0,2;1,1;0} \Bigg] ,\\
({\rm XVIII})+({\rm XIX})+({\rm XX})&=-\frac{4\tilde y_D{\tilde m}_M^R{\tilde m}_M^L \tilde G }{\pi}\Bigg[
3\left( 1-\frac{\eta_f}{6}\right) \left( 1  -18 {\tilde a} - 6\tilde b \right)
(1-{\tilde m}_M^L{\tilde m}_M^R){\mathcal I}_{0,2;1,1;0} +\nn
&\quad
+\left(1-\frac{\eta_\psi}{5}\right) \Big(1-{\tilde m}_M^R{\tilde m}_M^L \Big\{ 2+({\tilde m}_M^R)^2+({\tilde m}_M^L)^2 +{\tilde m}_M^R{\tilde m}_M^L\Big\}\Big){\mathcal I}_{0,1;2,2;0} \Bigg] ,\\
({\rm XXI})+
\cdots+({\rm XXIV})&=-\frac{6\tilde y_D \tilde \lambda_2 \tilde G}{5\pi}\Bigg[ \left(1-\frac{\eta_\psi}{6}\right)\Big(\left(1-(\tilde m_M^L)^2 \right){\mathcal I}_{0,1;2,0;1} + \left(1-(\tilde m_M^R)^2 \right){\mathcal I}_{0,1;0,2;1} \Big) +\nn
&\quad
+\frac{2}{3}\left\{ 4\left( 1-\frac{\eta_f}{7}\right) \left( 1  -9{\tilde a} -3\tilde b \right)+\left( 1-\frac{\eta_g}{7}\right) \left( 5-48 {\tilde a} -16\tilde b\right)  \right\} 
{\mathcal I}_{0,2;1,0;1} + \nn
&\quad
+\frac{2}{3}\left\{ 4\left( 1-\frac{\eta_f}{7}\right) \left( 1  -9{\tilde a} -3\tilde b \right)+\left( 1-\frac{\eta_g}{7}\right) \left( 5-48 {\tilde a} -16\tilde b\right)  \right\} {\mathcal I}_{0,2;0,1;1}  +\nn
&\quad
+2\left( 1-\frac{\eta_\phi}{7}\right) \Big({\mathcal I}_{0,1;1,0;2} +{\mathcal I}_{0,1;0,1;2}  \Big)
\Bigg] ,\\
({\rm XXV})+({\rm XXVI})+({\rm XXVII})&=\frac{8\tilde y_D \tilde \lambda_2 \tilde G}{\pi}\Bigg[\left(1-\frac{\eta_\psi}{5}\right) \Big(({\tilde m}_M^L)^2 {\mathcal I}_{0,1;2,0;1} +({\tilde m}_M^R)^2 {\mathcal I}_{0,1;0,2;1}\Big) +\nn
&\quad
+3\left(1 -\frac{\eta_f}{6}\right)\left(1 -9{\tilde a} -3 \tilde b \right)  \Big( ({\tilde m}_M^L)^2{\mathcal I}_{0,2;1,0;1}+({\tilde m}_M^R)^2{\mathcal I}_{0,2;0,1;1}  \Big) +\nn
&\quad
+\left(1-\frac{\eta_\phi}{6}\right) \Big(({\tilde m}_M^L)^2 {\mathcal I}_{0,1;1,0;2} +({\tilde m}_M^R)^2 {\mathcal I}_{0,1;0,1;2}\Big)
\Bigg] .
\end{align}
\end{subequations}
The diagrams of these contributions are shown in Fig.~\ref{Feynman diagrams of Yukawa beta functions}.

For the discussion presented section \ref{results_2} we need the contribution of Majorana masses to running of the scalar potential $V(\phi^2)$. Such a contribution comes exclusively from the fermionic loop corresponding to $\mathcal{T}_{(1/2)}$. Keeping only terms up to order $\mathcal{O}(\bar R)$ we find
\begin{align}
\mathcal{T}_{(1/2)} 
=&-\frac{k^4}{16\pi^2} \left( 1-\frac{\eta_\psi}{5} \right) \int d^4x\, \sqrt{\bar g}\,
\frac{\left(1+ \frac{(\tilde m_M^R)^2}{1+\tilde y_D^2\tilde \phi^2/k^2} \right)+ \left(1+ \frac{(\tilde m_M^L)^2}{1+\tilde y_D^2\tilde \phi^2/k^2}\right)}{\left(1+ \frac{(\tilde m_M^R)^2}{1+\tilde y_D^2\tilde \phi^2/k^2} \right)\left(1+ \frac{(\tilde m_M^L)^2}{1+\tilde y_D^2\tilde \phi^2/k^2} \right) +\left(1-\frac{\tilde m_M^R\tilde m_M^L}{1+\tilde y_D^2\tilde \phi^2/k^2} \right)^2 \tilde y_D^2 \tilde \phi^2 /k^2 }
 +
\nn
&+\frac{k^2}{96\pi^2} \left( 1-\frac{\eta_\psi}{3} \right) \int d^4x\, \sqrt{\bar g}
\,\left[
\frac{1}{1+ (\tilde m_M^R)^2}
+ \frac{1}{1+ (\tilde m_M^L)^2}
\right]\, {\bar R}
\,+\mathcal{O}(\bar R^2) \,.
\label{fermion loop effect on f(R)}
\end{align}
We kept the contribution proportional to $\bar R$ since this it contributes to the running of the Newton coupling, which we used in order to compute the fixed points in the appendix \ref{Fixed point analysis}.
The dots denote corrections to higher dimensional operators which are truncated in the present analysis such as $R^2$, $F(\phi)R$ and $F(\phi)R^2$.
Contributions to the scalar potential coming from graviton and scalar loops was evaluated in e.g.~\cite{Narain:2009fy,Oda:2015sma}.

\section{Fixed point analysis}\label{Fixed point analysis}

In this section we present the fixed point structure obtained from our truncation. The idea is to establish a benchmark analysis and investigate the possibility of non-trivial fixed points for the Majorana masses. In this sense, we restrict ourselves to the Einstein-Hilbert truncation for the gravitational sector. Further investigations are necessary in order to check the stability of our results by enlarging our truncation.
The results presented in this section were obtained for a maximally symmetric background.\footnote{
A complete and consistent computation of the full effective action must encode split Ward identities which take into account the background-independent nature of the theory, see, e.g., for some discussions on that \cite{Becker:2014qya,Denz:2016qks,Labus:2016lkh}.
}
In such a case we have
\begin{align}
\bar R_{\mu\nu\alpha\beta} = \frac{1}{12} (\bar g_{\mu\alpha} \bar g_{\nu\beta} - \bar g_{\mu\beta} \bar g_{\nu\alpha}) \bar R ,
\qquad\qquad
\bar R_{\mu\nu} = \frac{1}{4} \bar g_{\mu\nu} \bar R .
\end{align}
For the scalar potential we consider the $\phi^4$-truncation
\begin{align}
\tilde{V}(\tilde{\phi}^2) = k^{2}\, \tilde{\lambda}_{2} \, \tilde{\phi}^{2} + \tilde{\lambda}_{4} \, \tilde{\phi}^{4}.
\end{align}
In the present analysis we neglect any contributions coming from anomalous dimensions. In addition, the contribution of matter fluctuations on the running of the gravitational couplings was only considered by coming from the kinetic terms of the matter fields. These can be faced as additional approximations in our calculation and should be improved in future investigations.

Within this framework we found three classes of fixed points. Table \ref{Table_Fixed_Points} shows the numerical results for the fixed points, while in table \ref{Table_Critical_Exponents} we report the corresponding critical exponents. The first class correspond to the case of Gaussian matter. In fact, inspecting the beta functions corresponding to matter couplings it is not difficult to conclude that the subsystem fixed point equations is satisfied by Gaussian matter couplings, irrespective of the choice for the gravitational couplings. In this case, it is still necessary to investigate the back-reaction of Gaussian matter fields on the gravitational sector. Some systematic analysis in this direction has been done for Einstein-Hilbert truncation~\cite{Dona:2013qba} and $f(R)$-truncation~\cite{Alkofer:2018fxj}. In addition to the usual 2-relevant directions present in the Einstein-Hilbert truncation, there is an additional relevant direction associated with the coupling $\tilde{\lambda}_2$. 

The second class is characterized by one of the Majorana masses being interacting at the fixed point, while the remaining matter couplings vanish. Following the same reasoning discussed in the first class of fixed points, taking into account the subsystem of beta functions associated with the matter coupling, but one of the Majorana masses, it is not difficult to see that this sub-sector admits a Gaussian solution, irrespective of the fixed point values for the gravitational couplings and the other Majorana mass. Moreover, a fixed point configuration where only one of the Majorana masses is interacting, the Yukawa coupling is necessarily Gaussian\footnote{This statement is not valid if we include some additional symmetry in order to prevent one the Majorana masses.}. In comparison with the first class, we can observe the appearance of an additional relevant direction which can be associated with the interacting Majorana mass.

Finally, the third class of fixed points is characterized by interacting Majorana masses. Moreover, for the remaining matter couplings (Yukawa and scalar potential sectors) we found the usual Gaussian solution, which can be verified irrespective of the fixed point values for the gravitational couplings and the Majorana masses. In this case, both left- and right-handed masses exhibit the same values at the fixed point, which is consistent with the symmetry $\text{Left} \,\, \leftrightarrow \,\, \text{Right}$ present in our flow equation. In principle, it would also be possible to find fixed points with $\tilde{m}_M^L \neq \tilde{m}_M^R$, provided they come in pairs in order to preserve the aforementioned discrete symmetry. However, in our numerical analysis we found only fixed points candidates with equal left- and right-handed masses. As one can see, in comparison with the first class of fixed points, the present case has two additional relevant directions which can associated with the Majorana masses.

\begin{table}[htb!]
	\renewcommand\arraystretch{1.3}
	\begin{tabular}{| C | C | C | C | C | }\hline\hline
		\hspace*{1.5cm}   & \hspace*{.8cm} \tilde{\Lambda } \hspace*{.8cm} &
		\hspace*{.8cm} \tilde{G} \hspace*{.8cm}& \hspace*{.4cm}\tilde{m}_{R,(L)} \hspace*{.4cm}& \hspace*{.4cm} \tilde{m}_{L,(R)} \hspace*{.4cm}  \\ \hline
		1^{\textmd{st.}}  \textmd{class}  & -0.146252 & 3.20363 & 0 & 0  \\ 
		2^{\textmd{nd.}}  \textmd{class}  & -0.146252 & 3.20363 & 0.257486 & 0  \\
		3^{\textmd{rd.}}  \textmd{class}  & -0.146252 & 3.20363 & 0.257486 & 0.257486  \\
		\hline\hline
	\end{tabular}
	\caption{Fixed point structure obtained within our truncation. Those couplings which are not listed in this table vanish at the fixed point.}
	\label{Table_Fixed_Points}
\end{table}

\begin{table}[htb!]
	\renewcommand\arraystretch{1.3}
	\begin{tabular}{| C | C | C | C | C | C | C | C | }\hline\hline
		\hspace*{1.5cm} & \theta_1 & \theta_2 & \theta_3 & \theta_4 & \theta_5 & \theta_6 & \theta_7  \\ \hline
		1^{\textmd{st.}}  \textmd{class}  & \,2.92164\, & \,0.728222\, & \,0.235922\, & \,-1.76408\, & \,-1.0474\, & \,-0.047397\, & \,-0.047397\,  \\ 
		2^{\textmd{nd.}}  \textmd{class}  & \,2.92164\, & \,0.728222\, & \,0.235922\, & \,0.0848264\, & \,-1.76408\, & \,-1.0\, & \,-0.047397\, \\
		3^{\textmd{rd.}}  \textmd{class}  & \,2.92164\, & \,0.728222\, & \,0.235922\, & \,0.0848264\, & \,0.0848264\, & \,-1.76408\, & \,-0.915174\, \\
		\hline\hline
	\end{tabular}
	\caption{Critical exponents corresponding to the three classes of fixed point exhibited in table \ref{Table_Fixed_Points}.}
	\label{Table_Critical_Exponents}
\end{table}

\bibliography{majorana}
\end{document}